\documentclass[UTF8,prc,twocolumn,preprintnumbers,amsmath,amssymb]{revtex4-2}
\usepackage{CJK}
\usepackage{tipa}
\usepackage{graphicx}
\usepackage{float}
\usepackage{dcolumn}
\usepackage{bm}
\usepackage{multirow}
\usepackage{booktabs}
\usepackage{placeins}
\usepackage{textcomp}
\usepackage{color,xcolor}
\usepackage{comment}
\definecolor{darkred}{rgb}{139,0,0}

\usepackage[dvipdfm,
            pdfstartview=FitH,
            bookmarks=true,
            bookmarksnumbered=true,
            bookmarksopen=true,
            bookmarksopenlevel=2,
            colorlinks=true,
            pdfborder=001,
            linkcolor=blue,
            anchorcolor=green,
            urlcolor=blue,
            citecolor=blue
            ]{hyperref}

\begin{document}
\begin{CJK*}{UTF8}{song}
\title{Extension of the particle x-ray coincidence technique: The lifetimes and branching ratios apparatus}

\date{\today}

\author{L.~J.~Sun$^{1}$}
\email{sunli@frib.msu.edu}
\author{J.~Dopfer$^{2,1}$}
\author{A.~Adams$^{2,1}$}
\author{C.~Wrede$^{2,1}$}
\email{wrede@frib.msu.edu}
\author{A.~Banerjee$^{3,4}$}
\author{B.~A.~Brown$^{2,1}$}
\author{J.~Chen$^{1}$}
\author{E.~A.~M.~Jensen$^{5}$}
\author{R.~Mahajan$^{1}$}
\author{T.~Rauscher$^{6,7}$}
\author{C.~Sumithrarachchi$^{1}$}
\author{L.~E.~Weghorn$^{2,1}$}
\author{D.~Weisshaar$^{1}$}
\author{T.~Wheeler$^{2,1,8}$}

\affiliation{\footnotesize
$^1$Facility for Rare Isotope Beams, Michigan State University, East Lansing, Michigan 48824, USA\\
$^2$Department of Physics and Astronomy, Michigan State University, East Lansing, Michigan 48824, USA\\
$^3$Saha Institute of Nuclear Physics, Kolkata, West Bengal 700064, India\\
$^4$Homi Bhabha National Institute, Anushaktinagar, Mumbai 400094, India\\
$^5$Institut for Fysik \& Astronomi, Aarhus Universitet, Aarhus C 8000, Denmark\\
$^6$Department of Physics, University of Basel, 4056 Basel, Switzerland\\
$^7$Centre for Astrophysics Research, University of Hertfordshire, Hatfield AL10 9AB, UK\\
$^8$Department of Computational Mathematics, Science, and Engineering, Michigan State University, East Lansing, Michigan 48824, USA\\
{\color{white}{***********************************************************************************}}
}

\begin{abstract}
The particle x-ray coincidence technique (PXCT) was originally developed to measure average lifetimes in the $10^{-17}-10^{-15}$~s range for proton-unbound states populated by electron capture (EC). We have designed and built the Lifetimes and Branching Ratios Apparatus (LIBRA) to be used in the stopped-beam area at the Facility for Rare Isotope Beams that extends PXCT to measure lifetimes and decay branching ratios of resonances populated by EC/$\beta^+$ decay. The first application of LIBRA aims to obtain essential nuclear data from $^{60}$Ga EC/$\beta^+$ decay to constrain the thermonuclear rates of the $^{59}$Cu$(p,\gamma)^{60}$Zn and $^{59}$Cu$(p,\alpha)^{56}$Ni reactions, and in turn, the strength of the NiCu nucleosynthesis cycle, which is predicted to significantly impact the modeling of type I x-ray burst light curves and the composition of the burst ashes. Detailed theoretical calculations, Monte Carlo simulations, and performance tests with radioactive sources have been conducted to validate the feasibility of employing LIBRA for the $^{60}$Ga experiment. LIBRA can be utilized to measure most essential ingredients needed for charged-particle reaction rate calculations in a single experiment, in the absence of direct measurements, which are often impractical for radioactive reactants.
\end{abstract}
\maketitle
\end{CJK*}

\section{Introduction}
Direct measurements of charged-particle thermonuclear reaction rates are challenging, especially when radioactive reactants are involved. Small cross sections at stellar energies, limited beam intensities, target degradation under bombardment, and low signal-to-background ratios may render a direct measurement infeasible. Successful direct measurements at astrophysical energies have been achieved only in a few instances~\cite{Aliotta_JPG2022}. Consequently, direct measurements of thermonuclear reaction rates are often performed at higher energies and then extrapolated to stellar energies with the aid of nuclear theories. Alternatively, various indirect approaches, such as elastic scattering, transfer reactions, surrogate reactions, charge-exchange reactions, Coulomb dissociation, in-beam $\gamma$-ray spectroscopy, and $\beta$-decay spectroscopy have also played a major role in achieving a comprehensive understanding of thermonuclear reactions~\cite{Tribble_RPP2014,Brune_ARNPS2015,Hammache_FP2021}. However, these methods typically yield only a fraction of the necessary nuclear data, and results from multiple experiments still need to be pieced together to infer the reaction rates of interest~\cite{Angulo_NPA1999}.

Thermonuclear charged-particle reactions are often dominated by narrow and isolated resonances if the level density of the compound nucleus in the Gamow window is not too high. The resonant reaction rate $N_A\langle\sigma\nu\rangle_r$ can be calculated using the well-known expression~\cite{Rolfs_1988,Iliadis_2015}

\begin{equation}
\begin{split}
N_A\langle\sigma\nu\rangle_r=1.5394\times10^{11}(\mu T_9)^{-3/2}\times\omega\gamma\\
\times\exp\left(-\frac{11.605E_r}{T_9}\right)(\mathrm{cm^3s^{-1}mol^{-1}}),
\end{split}
\label{eq:ReactionRate}
\end{equation}

where $N_A$ denotes the Avogadro constant and $\langle\sigma\nu\rangle_r$ is the velocity-averaged product of the cross section and relative velocity. $\mu=A_pA_T/(A_p+A_T)$ is the reduced mass in atomic mass units, with $A_p$ and $A_T$ as the mass numbers of the incoming particle and the target nucleus, respectively. $E_r$ is the resonance energy in the center-of-mass system in units of MeV. $T_9$ is the temperature in units of giga kelvin (GK), and $\omega\gamma$ is the resonance strength in units of MeV. For a $(p,\gamma)$ resonance,

\begin{equation}
\omega\gamma=\frac{2J_r+1}{(2J_p+1)(2J_{T}+1)}\frac{\Gamma_p\Gamma_\gamma}{\Gamma_\mathrm{tot}},
\end{equation}

where $J_r$ is the spin of the resonance, $J_p=1/2$ is the spin of proton, and $J_T$ is the spin of the ground state of the target nucleus. The total decay width $\Gamma_{\mathrm{tot}}$ of the resonance is the sum of the partial widths for open decay channels, typically including proton width ($\Gamma_p$), $\gamma$ width ($\Gamma_\gamma$), and $\alpha$ width ($\Gamma_\alpha$). Equivalently, the resonance strength can be constructed by combining the proton branching ratio $B_p=\Gamma_p/\Gamma_{\mathrm{tot}}$, the $\gamma$-ray branching ratio $B_\gamma=\Gamma_\gamma/\Gamma_{\mathrm{tot}}$, and the lifetime $\tau$ using the following expression:

\begin{equation}
\omega\gamma=\frac{2J_r+1}{(2J_p+1)(2J_{T}+1)}B_pB_\gamma\frac{\hbar}{\tau},
\label{eq:ResonanceStrength}
\end{equation}

where $\hbar$ is the reduced Planck constant. These relations are also applicable to a $(p,\alpha)$ resonance by replacing the terms $\Gamma_\gamma$ and $B_\gamma$ with $\Gamma_\alpha$ and $B_\alpha$, respectively. Therefore, important quantities to determine the reaction rates include the resonance energies, the spins, the proton-, $\gamma$-, and $\alpha$-decay branching ratios, and the lifetimes of the resonances.

In cases where the level density of resonances in the compound nucleus is sufficiently high to justify a statistical treatment, the $(p,\gamma)$ reaction cross section $\sigma_{p\gamma}$ can be estimated within the Hauser-Feshbach statistical model framework:

\begin{equation}
\sigma_{p\gamma} = \frac{\pi \hbar^2}{2\mu E_r(2J_p + 1)(2J_T + 1)} \sum_{J,\pi} (2J + 1) \frac{T_p^{J^\pi} T_\gamma^{J^\pi}}{\sum_k T_k^{J^\pi}},
\label{eq:HauserFeshbach}
\end{equation}

where $\mu$ is the reduced mass, $E_r$, $J$, and $\pi$ are the energy, spin, and parity of the resonance in the compound nucleus, and $J_p$ and $J_T$ are the spins of the proton and the state in the target nucleus, respectively. $T_p^{J^\pi}$ and $T_\gamma^{J^\pi}$ are the transmission coefficients for the proton and $\gamma$ channels, respectively, of a given resonance with $J,\pi$ at $E_r$. $\sum_k T_k^{J^\pi}$ represents the sum of the transmission coefficients over all possible decay channels $k$ of the resonance, including proton, $\alpha$, and $\gamma$ emissions. The statistical model assumes that a large number of resonances with all spins and parities are available at each energy, and hence, Eq.~\eqref{eq:HauserFeshbach} includes the summation over $J,\pi$. Transmission coefficients quantify the probability of particles or photons transmitting through the nuclear potential barrier, and are related to the average decay widths through the number of resonances with $J,\pi$ per energy interval, i.e., the nuclear level density. Therefore, obtaining average resonance properties, including decay branching ratios and lifetimes (or total widths) of excited states in the compound nucleus, is valuable for calculating reaction rates within the statistical model~\cite{Rauscher_ADNDT2000,Rauscher_2020}.

In this paper, we introduce the Lifetimes and Branching Ratios Apparatus (LIBRA) that applies and extends the Particle X-ray Coincidence Technique (PXCT)~\cite{Hardy_PRL1976} to measure various essential ingredients for thermonuclear reaction rate calculations in a single experiment, potentially reducing uncertainties associated with combining quantities from separate experiments. We provide a comprehensive description of the experimental setup and its performance tests, demonstrating the feasibility of employing LIBRA in a case study to address the question of NiCu cycling in Type I X-ray bursts (XRBs).

\section{Case Study: NiCu Cycle in XRBs}
Type I XRBs are the most frequent type of thermonuclear stellar explosions in the Galaxy~\cite{Jose_2016}. They are powered by thermonuclear runaways in hydrogen- and/or helium-rich material accreted onto the surface of a neutron star in a low-mass X-ray binary system. The main nuclear reaction flow in the XRB is driven towards the proton drip line and to higher masses via a series of particle-induced reactions and $\beta^+$ decays. Accurate modeling of energy production and nucleosynthesis in XRBs requires precise nuclear physics inputs, including $\beta$ decay rates, nuclear masses, and nuclear reaction rates of proton-rich rare isotopes along the path of the rapid proton ($rp$) capture process. Our understanding of XRBs has greatly expanded thanks to decades of work, yet many open questions remain~\cite{Schatz_NPA2006,Parikh_PPNP2013}.

\begin{figure}[hbp!]
\begin{center}
\includegraphics[width=8cm]{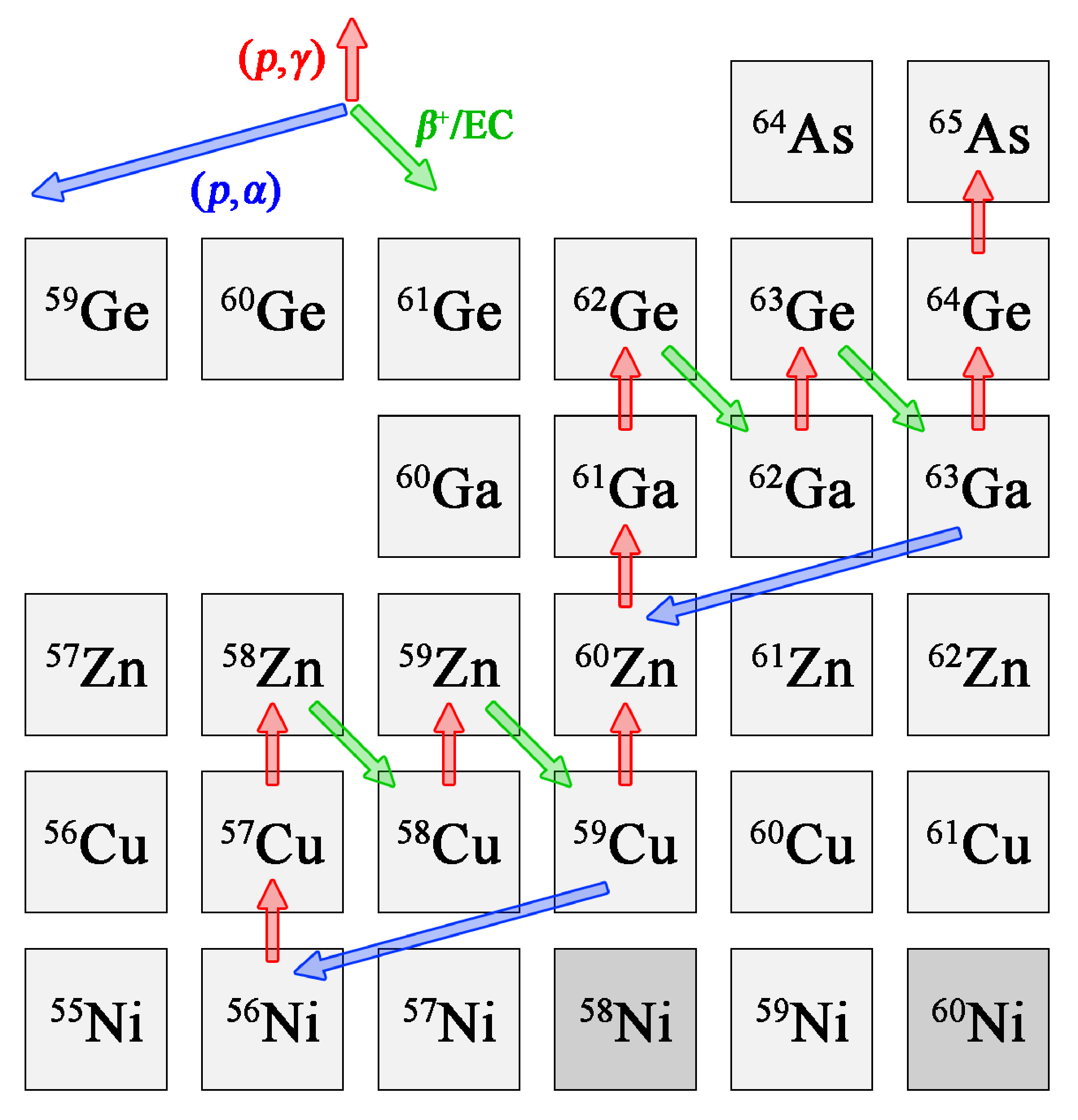}
\caption{\label{NiCu_Cycle}Portion of the $rp$-process reaction sequence featuring the NiCu cycle and ZnGa cycle one $\alpha$ cluster above. $^{58}$Ni and $^{60}$Ni (darker gray) are stable isotopes.}
\end{center}
\end{figure}

As illustrated in Fig.~\ref{NiCu_Cycle}, under XRB conditions, the $rp$ process beyond the waiting point $^{56}$Ni may be affected by several cycles, such as the NiCu cycle. The ratio of $^{59}$Cu$(p,\alpha)^{56}$Ni to $^{59}$Cu$(p,\gamma)^{60}$Zn rate could lead to the formation of a NiCu cycle, returning the reaction flux to $^{56}$Ni, strongly impeding the synthesis of heavier nuclei and affecting the XRB observables~\cite{Van_Wormer_APJ1994}. Currently, both rates recommended by REACLIB~\cite{Cyburt_APJS2010} are calculated by the Hauser-Feshbach statistical model~\cite{Rauscher_ADNDT2000,Rauscher_ADNDT2001}. The variations in these rates have been identified as having a significant impact on the modeling of XRB light curves and the composition of the burst ashes~\cite{Parikh_APJS2008,Cyburt_APJ2016,Meisel_APJ2019}. At higher temperatures ($\approx$3~GK), the competition between $^{59}$Cu$(p,\gamma)^{60}$Zn and $^{59}$Cu$(p,\alpha)^{56}$Ni reactions is also found to significantly impact the $\nu p$-process nucleosynthesis in core-collapse supernovae~\cite{Frohlich_PRL2006,Arcones_APJ2012,Hermansen_APJ2020}.

It is challenging to directly measure these two reactions at astrophysical energies because the predicted cross sections are too small, and intense low-energy radioactive $^{59}$Cu beams are not currently available. A $^{59}$Cu$(p,\alpha)^{56}$Ni reaction measurement using a $^{59}$Cu beam with an intensity of $3.6\times10^{3}$~particle per second (pps) on a cryogenic solid H$_2$ target at an excitation energy of $E_x(^{60}\mathrm{Zn})=11.1$~MeV found that $^{59}$Cu$(p,\alpha)$ proceeds predominantly to the $^{56}$Ni ground state, and standard statistical model calculations overestimate the cross section by a factor of 1.6$-$4~\cite{Randhawa_PRC2021}. In a $^{58}$Ni$(^3$He$,n)^{60}$Zn reaction measurement~\cite{Soltesz_PRC2021}, the nuclear level density of $^{60}$Zn was extracted from the neutron evaporation spectrum. At an excitation energy of 6~MeV, the level density was estimated to be $\approx$18~MeV$^{-1}$. Taking into account the spin and parity range relevant to $\ell=0,1$ proton captures~(Table~\ref{Selection_rule}), it was concluded that the level density could be too low to accurately apply the Hauser-Feshbach statistical model. Kim \textit{et al}.~\cite{Kim_APJ2022} evaluated available experimental data on $^{60}$Zn resonances, supplemented with theoretical calculations. They found the $^{59}$Cu$(p,\alpha)^{56}$Ni reaction rate to be lower than the REACLIB rate~\cite{Cyburt_APJS2010} at XRB temperatures, implying a weaker NiCu cycle strength than previously estimated~\cite{Parikh_APJS2008,Cyburt_APJ2016,Meisel_APJ2019}.

There are many ongoing efforts to address this problem both directly and indirectly:

1) $^{56}$Ni$(\alpha,p)^{59}$Cu reaction cross section measurement using a $^{56}$Ni beam of $3\times10^{3}$~pps on a He jet target at $E_x(^{60}\mathrm{Zn})=11.1, 11.7, 12.6$~MeV with the Jet Experiments in Nuclear Structure and Astrophysics setup~\cite{E18039_JENSA};

2) $^{59}$Cu$(p,\alpha)^{56}$Ni reaction cross section measurement using a $^{59}$Cu beam of $2\times10^{4}$~pps on CH$_4$ gas within the Multi-Sampling Ionization Chamber at $E_x(^{60}\mathrm{Zn})=7.3-11.0$~MeV ~\cite{E21026_MUSIC};

3) $^{59}$Cu$(p,\alpha)^{56}$Ni reaction cross section measurement using a $^{59}$Cu beam of $5.5\times10^{5}$~pps on a CH$_2$ target at $E_x(^{60}\mathrm{Zn})=8.3-10.1$~MeV with circular double-sided silicon detectors~\cite{Lederer_CERN2015};

4) $^{60}$Zn $\gamma$-ray spectroscopy via the $^{59}$Cu$(d,n)^{60}$Zn transfer reaction using Gamma-Ray Energy Tracking In-beam Nuclear Array~\cite{E21014_GRETINA};

5) $^{58}$Ni$(^3$He$,n)^{60}$Zn reaction using a $^3$He beam on a $^{58}$Ni target and measuring neutron angular distributions using liquid scintillators to determine the spins and parities of $^{60}$Zn states~\cite{Okamura_Thesis2024};

6) $^{59}$Cu$(^3$He$,d)^{60}$Zn reaction using a $^{59}$Cu beam to populate $^{60}$Zn states and to measure the decay branching ratios~\cite{Furuno_PC2024};

7) $^{60}$Ga $\beta$-delayed $\gamma$ decay total absorption spectroscopy using the Summing NaI detector to determine the $\beta$-decay strength distribution and $\gamma$-ray strength functions~\cite{E17009_SUN};

8) $^{60}$Ga decay using the Gaseous Detector with Germanium Tagging II to discover resonances and to measure decay branching ratios~\cite{E23035_GADGET}.

To date, experimental constraints on the $^{59}$Cu$(p,\gamma)^{60}$Zn and $^{59}$Cu$(p,\alpha)^{56}$Ni are still scarce and preclude a robust understanding of their astrophysical impacts.

Table~\ref{Selection_rule} summarizes the spins and parities of $^{59}\text{Cu}+p$ resonances in $^{60}$Zn. We also include captures on the first excited state of $^{59}$Cu, which can be thermally populated at XRB temperatures~\cite{Rauscher_ADNDT2000}. Only positive parity states associated with $\ell=1$ proton captures are accessible via allowed $^{60}$Ga $\beta$ transitions, also indicating a lower density of levels populated in the $\beta$ decay than in the previous $^{58}$Ni$(^3$He$,n)^{60}$Zn reaction measurement~\cite{Soltesz_PRC2021}.

\begin{table}[htbp!]
\caption{\label{Selection_rule}Properties of $^{60}$Zn states populated via proton captures on the $3/2^-$ $^{59}$Cu ground state and the $1/2^-$ $^{59}$Cu first excited state, and the allowed $\beta$ transitions of the $2^+$ $^{60}$Ga ground state.}
\begin{center}
\renewcommand{\arraystretch}{1.1}  
\begin{ruledtabular}
\begin{tabular}{cc}
Population & $^{60}$Zn states \\
\hline
$\ell=0$ $p$ on $3/2^-$ & $1^-$, $2^-$ \\
$\ell=1$ $p$ on $3/2^-$ & $0^+$, $1^+$, $2^+$, $3^+$ \\
$\ell=2$ $p$ on $3/2^-$ & $0^-$, $1^-$, $2^-$, $3^-$, $4^-$ \\
$\ell=0$ $p$ on $1/2^-$ & $0^-$, $1^-$ \\
$\ell=1$ $p$ on $1/2^-$ & $0^+$, $1^+$, $2^+$ \\
$\ell=2$ $p$ on $1/2^-$ & $1^-$, $2^-$, $3^-$ \\
$\beta$ from $2^+$ & $1^+$, $2^+$, $3^+$\\
\end{tabular}
\end{ruledtabular}
\end{center}
\end{table}

The Gamow energies and windows for the $^{59}$Cu$(p,\gamma)^{60}$Zn and $^{59}$Cu$(p,\alpha)^{56}$Ni reactions shown in Table~\ref{Gamow} are calculated from a numerical study of the relevant energy ranges for astrophysical reaction rates~\cite{Rauscher_PRC2010}. For XRB, the most relevant Gamow windows are those calculated at temperatures of 0.5$-$1.5~GK~\cite{Iliadis_2015}. Combined with the proton-separation energy of $^{60}$Zn $S_{p}(^{60}$Zn$)=5105.0(4)$~keV~\cite{Wang_CPC2021} and $\alpha$-separation energy of $^{60}$Zn $S_{\alpha}(^{60}$Zn$)=2691.7(5)$~keV~\cite{Wang_CPC2021}, $^{60}$Zn resonances of interest are energetically accessible in $^{60}$Ga $\beta$ decay owing to the large $Q_{\mathrm{EC}}(^{60}$Ga$)=14160(15)$~keV~\cite{Orrigo_PRC2021,Paul_PRC2021,Wang_PRL2023}.

\begin{table*}[htbp!]
\caption{\label{Gamow} Gamow windows $\widetilde E_{\rm hi}-\widetilde \Delta\leq E\leq\widetilde E_{\rm hi}$ and Gamow peaks $\widetilde E_0$ for the $^{59}$Cu$(p,\gamma)^{60}$Zn and $^{59}$Cu$(p,\alpha)^{56}$Ni reactions at a temperature $T$~\cite{Rauscher_PRC2010}.}
\begin{center}
\renewcommand{\arraystretch}{1.1}  
\begin{ruledtabular}
\begin{tabular}{ccccccc}
  & \multicolumn{3}{c}{$^{59}$Cu$(p,\gamma)^{60}$Zn} & \multicolumn{3}{c}{$^{59}$Cu$(p,\alpha)^{56}$Ni} \\
  \cline{2-4} \cline{5-7}
$T$~(GK) & $\widetilde E_{\rm hi}-\widetilde \Delta$~(MeV) & $\widetilde E_0$~(MeV)  & $\widetilde E_{\rm hi}$~(MeV) & $\widetilde E_{\rm hi}-\widetilde \Delta$~(MeV) & $\widetilde E_0$~(MeV)  & $\widetilde E_{\rm hi}$~(MeV) \\
  \hline
0.5	&	0.51	&	0.71	&	0.92	&	0.55	&	0.74	&	0.98	\\
1.0	&	0.67	&	0.91	&	1.26	&	0.73	&	1.01	&	1.48	\\
1.5	&	0.75	&	1.01	&	1.57	&	0.87	&	1.27	&	2.11	\\
2.0	&	0.82	&	1.14	&	1.83	&	1.01	&	1.74	&	2.80	\\
2.5	&	0.85	&	1.40	&	2.05	&	1.24	&	2.19	&	3.52	\\
3.0	&	0.89	&	1.49	&	2.26	&	1.51	&	2.66	&	4.16	\\
\end{tabular}
\end{ruledtabular}
\end{center}
\end{table*}

\begin{figure*}[htbp!]
\begin{center}
\includegraphics[width=17.2cm]{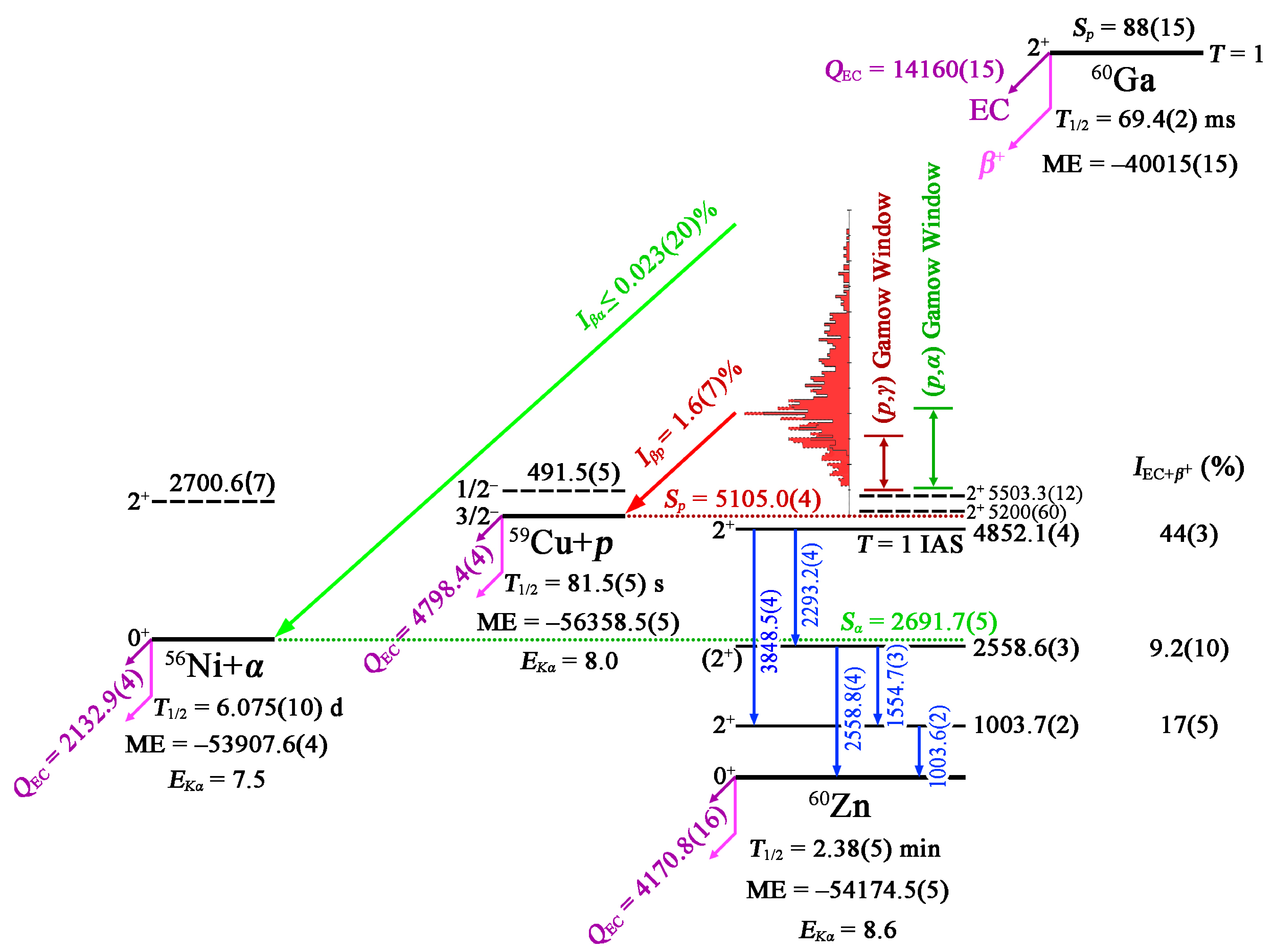}
\caption{\label{Scheme_60Ga} Evaluated decay scheme of $^{60}$Ga. All energies are given in units of keV. The mass excesses, $Q_{\mathrm{EC}}$ values, and particle separation energies of $^{56}$Ni, $^{59}$Cu, and $^{60}$Zn are from AME2020~\cite{Wang_CPC2021}, while for $^{60}$Ga the data are evaluated based on Refs.~\cite{Orrigo_PRC2021,Paul_PRC2021,Wang_PRL2023}. The half-lives of $^{56}$Ni, $^{59}$Cu, and $^{60}$Zn are from ENSDF evaluations~\cite{Huo_NDS2011_A56,Basunia_NDS2018_A59,Browne_NDS2013_A60}, respectively. The half-life of $^{60}$Ga is evaluated based on Refs.~\cite{Mazzocchi_EPJA2001,LopezJimenez_PRC2002,Kucuk_EPJA2017,Goigoux_Thesis2017,Giovinazzo_APPB2020,Orrigo_PRC2021}. All spins and parities are adopted from ENSDF evaluations~\cite{Huo_NDS2011_A56,Basunia_NDS2018_A59,Browne_NDS2013_A60}, with the 4852-keV state in $^{60}$Zn revised from $(2^+)$ to $2^+$ based on the unambiguous $T=1$ isobaric analog state argument~\cite{Orrigo_PRC2021,Mazzocchi_EPJA2001}. The $\gamma$-ray energies, excitation energies, and $\beta$ feedings of $^{60}$Zn states are evaluated~\cite{Chen_GLSC} based on all available measurements~\cite{Orrigo_PRC2021,Mazzocchi_EPJA2001,Kamermans_PRC1974,Svensson_Thesis1998}. The proton spectrum and $\beta$-delayed proton and $\alpha$ branchings are adopted from Ref.~\cite{Mazzocchi_EPJA2001}. Dashed lines represent two $2^+$ resonances in $^{60}$Zn and the first excited states of $^{56}$Ni and $^{59}$Cu, which are expected to be accessible but have not yet been observed in $^{60}$Ga decay. Double-headed arrows indicate the Gamow windows for the $^{59}$Cu$(p,\gamma)^{60}$Zn and $^{59}$Cu$(p,\alpha)^{56}$Ni reactions at 0.5$-$1.5~GK (Table~\ref{Gamow}). }
\end{center}
\end{figure*}

In the first $^{60}$Ga decay study, Mazzocchi \textit{et al}. observed 802 protons and reported a total $^{60}$Ga $\beta p$ intensity of $I_p=1.6(7)$\% and an upper limit for $\beta\alpha$ intensity $I_\alpha\le0.023(20)$\%. They also observed five $^{60}$Ga$(\beta\gamma)$ transitions through three $^{60}$Zn bound states~\cite{Mazzocchi_EPJA2001}. Orrigo \textit{et al}.~\cite{Orrigo_PRC2021} confirmed these five $\beta\gamma$ transitions and the three proton-bound states, and reported 24 new $\beta\gamma$ transitions that are correlated with $^{60}$Ga implants. However, these new transitions were not placed in the decay scheme, nor were $\beta$-feeding intensities provided. Individual proton peaks were not resolved in either work~\cite{Mazzocchi_EPJA2001,Orrigo_PRC2021,Goigoux_Thesis2017}. The Evaluated Nuclear Structure Data File (ENSDF) for mass 60 is 12~years old, and we present an up-to-date $^{60}$Ga decay data evaluation to facilitate an accurate understanding of the pertinent nuclear structure properties (Fig.~\ref{Scheme_60Ga}). However, the evaluated decay scheme remains incomplete, with substantial unplaced $\beta\gamma$ intensities. The five $\gamma$ intensities reported by both studies~\cite{Orrigo_PRC2021,Mazzocchi_EPJA2001} are in good agreement, and if we deduce the corresponding $\beta$-feeding intensities, unplaced $\beta\gamma$ transitions likely account for $>$20\% of total $\beta$-feeding intensities. A recent $^{60}$Ga total absorption $\gamma$-ray spectroscopy experiment observed $\beta$-feeding intensities above the $^{60}$Zn proton separation energy~\cite{E17009_SUN}, indicating the need for further studies.

High-statistics $^{60}$Ga $\beta$ decay measurements with proton, $\alpha$, and $\gamma$-ray coincidences will allow for the construction of a more comprehensive decay scheme, including the proton/$\alpha$-emitting states in $^{60}$Zn to the ground and excited states of $^{59}$Cu/$^{56}$Ni. This will provide crucial insights into the entrance and exit channels of the thermonuclear $^{59}$Cu$(p,\gamma)^{60}$Zn and $^{59}$Cu$(p,\alpha)^{56}$Ni reactions. Although $\beta$-decay spectroscopy has proven to be a powerful method for obtaining decay branching ratios, it typically does not yield lifetimes or widths of resonances~\cite{Budner_PRL2022,Sun_PLB2020}. Therefore, incorporating lifetime measurement capabilities into $\beta$-decay spectroscopy would be a significant advancement, allowing most essential information to be gathered in a single experiment.

\section{Particle X-ray Coincidence Technique}
\begin{figure}
\begin{center}
\includegraphics[width=8.5cm]{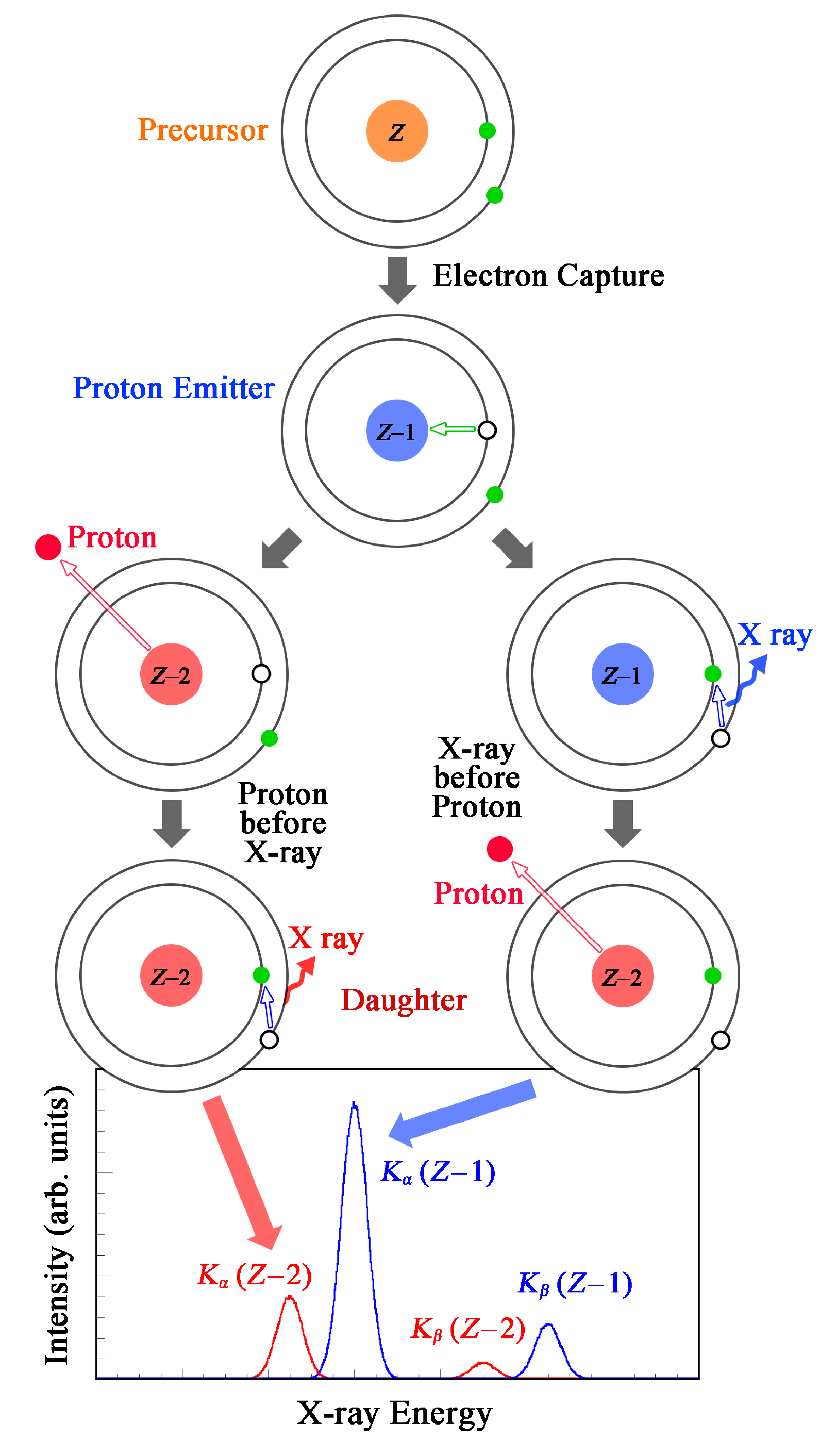}
\caption{\label{Cartoon} Schematic illustrating electron capture to a proton-unbound excited state and the Particle-X-ray Coincidence Technique.}
\end{center}
\end{figure}

In the 1970s, PXCT was introduced and applied to measure the average lifetimes of proton-unbound states in $^{69}$As populated by the electron capture (EC) of $^{69}$Se~\cite{Hardy_PRL1976}. The principle of PXCT is illustrated in Fig.~\ref{Cartoon}. In the process of an EC-delayed proton emission, a proton-rich precursor with an atomic number of $Z$ decays by EC to the proton emitter $(Z-1)$. Once EC occurs, a proton-unbound nuclear state and an atomic shell vacancy are created simultaneously. The vacancy is primarily created in the $K$ shell. An electron from an outer shell fills the $K$ shell vacancy and may yield X-ray photons corresponding to the binding energy difference between the outer and the $K$ shells. Meanwhile, the proton-unbound state with a comparable lifetime $\tau_{p-\mathrm{emit}}$ emits a proton to a state of the daughter $(Z-2)$. If the proton is emitted before the X-ray emission, then the X-ray energy will be characteristic of the daughter $(Z-2)$. If the proton is emitted after the X-ray emission, then the X-ray energy will be characteristic of the proton emitter $(Z-1)$. By measuring X rays in coincidence with protons, the relative intensities of the $(Z-1)$ and $(Z-2)$ X-ray peaks, primarily $I_{K_\alpha(Z-1)}/I_{K_\alpha(Z-2)}$, can be used to establish the relationship between the nuclear and atomic lifetimes:

\begin{equation}
\frac{\tau_{p-\mathrm{emit}}}{\tau_{K\mathrm{shell}(Z-1)}}=\frac{\Gamma_{K\mathrm{shell}(Z-1)}}{\Gamma_{p-\mathrm{emit}}}=\frac{I_{K_\alpha(Z-1)}}{I_{K_\alpha(Z-2)}},
\end{equation}

where the level widths $\Gamma_{K\mathrm{shell}}$ and $\Gamma_{p-\mathrm{emit}}$ are the equivalent of $\hbar/\tau_{K\mathrm{shell}}$ and $\hbar/\tau_{p-\mathrm{emit}}$, respectively, as they both follow the exponential decay law. The lifetimes of proton-emitting states can be determined by measuring X-ray intensity ratios combined with known atomic $K$-shell vacancy lifetimes, ranging from $1.1\times10^{-14}$~s for C $(Z=6)$ down to $5\times10^{-18}$~s for Fm $(Z=100)$~\cite{Perkins_EADL1991,Bambynek_RMP1972,Scofield_1975,Koziok_PRA2014,Campbell_ADNDT2001}. This also defines the PXCT applicable lifetime range, where conventional approaches are limited~\cite{Nolan_RPP1979,Massa_RNC1982}. The preceding discussion is also generalizable to EC-delayed $\alpha$-particle emission, where the proton-decay daughter $(Z-2)$ is replaced by $\alpha$-decay daughter $(Z-3)$. Another decay channel is EC-delayed $\gamma$-ray emission, which can occur either before or after the filling of atomic shell vacancies. However, the resulting X rays are always characteristic of $(Z-1)$ and are therefore insensitive for determining nuclear lifetimes.

So far, PXCT has been applied in the decay studies of six nuclei, as summarized in Table~\ref{PXCT_history}. A variation of PXCT has also been applied in $p+^{112}$Sn~\cite{Rohl_PRL1979,Rohl_NPA1981} and $p+^{106}$Cd~\cite{Chemin_NPA1979} inelastic scattering measurements at 10 and 12~MeV proton incident energies. By measuring X-rays in coincidence with inelastically and elastically scattered protons, the lifetimes of compound nuclear states can be deduced. However, an uncertain factor in this reaction process is the ionization probability of the $K$ shell during the incoming and outgoing parts of the collision~\cite{Amundsen_PRC1986,Heinz_RPP1987}. In the following discussion, we focus on the application of PXCT to EC decay.

In the previous PXCT studies, the proton energy distribution and the X-ray count ratios as a function of coincident proton energies are the most important observables. By tuning the statistical model parameters to reproduce the experimental observables, the model parameters, such as average proton partial widths ($\bigl \langle \Gamma_p\bigr \rangle$), average $\gamma$-ray partial widths ($\bigl \langle \Gamma_\gamma\bigr \rangle$), and level densities ($\rho$), can be constrained~\cite{Hardy_NPA1981,Macdonald_NPA1977,Asboe-Hansen_NPA1981,Giovinazzo_NPA2000}.

The initial $\rho$ is typically estimated using the back-shifted Fermi gas model~\cite{Gilbert_CJP1965,Dilg_NPA1973}. The initial $\bigl \langle \Gamma_\gamma\bigr \rangle$ is calculated using $\gamma$-ray strength functions~\cite{Hardy_PLB1982}, which can be parametrized by Lorentzian fits to giant resonance cross-section data~\cite{Berman_RMP1975,Johnson_PRC1977}. The initial $\bigl \langle \Gamma_p\bigr \rangle$ is calculated using

\begin{equation}
\bigl \langle \Gamma_p\bigr \rangle=\frac{\sum_{\ell} T_\ell(E_p)}{2\pi\rho(E_x,J,\pi)},
\end{equation}

where $T_\ell(E_p)$ is the transmission coefficient for protons with energy $E_p$ and angular momentum $\ell$, and $\rho(E_x,J,\pi)$ is the level density with spin and parity $J,\pi$ at excitation energy $E_x$. $T_\ell(E_p)$ is typically calculated using the optical model. The PXCT experimentally constrained particle transmission coefficients will help benchmark the local optical model potentials in this mass region~\cite{Avrigeanu_PRC2022}.
 








\begin{table*}\footnotesize
\caption{\label{PXCT_history}Properties of all nuclei that have been measured with PXCT. Columns 1$-$7 present the EC/$\beta^+$ decay, the half-life ($T_{1/2}$) of the precursor, the $\beta$-decay energy ($Q_{\mathrm{EC}}$), the proton-separation energy of the EC/$\beta^+$-decay daughter ($S_p$), the total intensity of EC/$\beta^+$-delayed protons ($I_p$), the primary X-ray energies that need to be distinguished, the known lifetime of the $K$-shell vacancy, and the lifetime range of proton-emitting states of the EC/$\beta^+$-decay daughter obtained in each study, respectively. The two $K_\alpha$ energies listed in column 6 correspond to the proton-emission daughter $(Z-2)$ and proton emitter $(Z-1)$, respectively, which are adopted from Ref.~\cite{Newville_XrayDB2023} rounded to the nearest 0.1 keV. The last two rows list the properties of $^{60}$Ga and $^{64}$As for comparison.}
\begin{center}
\renewcommand{\arraystretch}{1.2}  
\begin{ruledtabular}
\begin{tabular}{cccccccc}
EC/$\beta^+$-decay & $T_{1/2}$~(s) & $Q_{\mathrm{EC}}$~(keV)~\cite{Wang_CPC2021} & $S_p$~(keV)~\cite{Wang_CPC2021} & $I_p$~(\%) & $E_{K\alpha}$~(keV)~\cite{Newville_XrayDB2023} & $\tau_{K\mathrm{shell}}$~(fs)~\cite{Campbell_ADNDT2001} & $\tau_{p-\mathrm{emit}}$~(fs) \\
\hline
$^{65}$Ge$\rightarrow$$^{65}$Ga & 30.9(5)~\cite{Browne_NDS2010_A65} & 6179.3(23) & 3942.4(6) & 0.011(3)~\cite{Giovinazzo_NPA2000,Hardy_PLB1976,Vierinen_NPA1987} & 8.6, 9.2 & 0.374 & $\approx$1.7~\cite{Hardy_NPA1981} \\

$^{69}$Se$\rightarrow$$^{69}$As & 27.4(2)~\cite{Nesaraja_NDS2014_A69} & 6680(30) & 3420(30) & 0.052(10)~\cite{Macdonald_NPA1977,Dessagne_PRC1988} & 9.9, 10.5 & 0.315 & 0.3$-$3.3~\cite{Hardy_PRL1976,Giovinazzo_NPA2000,Macdonald_NPA1977} \\

$^{73}$Kr$\rightarrow$$^{73}$Br & 27.3(10)~\cite{Singh_NDS2019_A73} & 7094(9) & 3067(7) & 0.47(22)~\cite{Hornshoj_NPA1972,Miehe_EPJA1999} & 11.2, 11.9 & 0.264 & 0.3$-$2.7 ~\cite{Giovinazzo_NPA2000,Asboe-Hansen_PLB1978,Asboe-Hansen_NPA1981}\\

$^{77}$Sr$\rightarrow$$^{77}$Rb & 9.0(2)~\cite{Singh_NDS2012_A77} & 7027(8) & 3106(4) & 0.08(3)~\cite{Giovinazzo_NPA2000,Hardy_PLB1976} & 12.6, 13.4 & 0.222 & $\approx$1.5~\cite{Giovinazzo_NPA2000} \\

$^{113}$Xe$\rightarrow$$^{113}$I & 2.74(8)~\cite{Blachot_NDS2010_A113} & 8916(11) & 841(12) & 7(4)~\cite{Janas_EPJA2005b} & 27.5, 28.6 & 0.062 & 0.3$-$2.9~\cite{Janas_EPJA2005b} \\

$^{117}$Ba$\rightarrow$$^{117}$Cs & 1.75(7)~\cite{Blachot_NDS2002_A117} & 9040(260) & 740(60) &16(3)~\cite{Janas_EPJA2005a} & 29.8, 31.0 & 0.054 & $>$4.7~\cite{Janas_EPJA2005a} \\
\hline
$^{60}$Ga$\rightarrow$$^{60}$Zn & 0.0694(3)\footnotemark[1] & 14160(15)\footnotemark[1] & 5105.0(4) & 1.6(7)~\cite{Mazzocchi_EPJA2001} & 8.0, 8.6 & 0.406 &  \\

$^{64}$As$\rightarrow$$^{64}$Ge & 0.0690(14)~\cite{Singh_NDS2021_A64} & 14606(110)\footnotemark[2] & 5057(4) & unreported~\cite{Rubio_JPCS2019} & 9.2, 9.9 & 0.343 &  \\
\end{tabular}
\end{ruledtabular}
\footnotetext[1]{See Fig.~\ref{Scheme_60Ga} for evaluation details.}
\footnotetext[2]{Deduced based on $^{64}$As mass~\cite{Zhou_NP2023} and $^{64}$Ge mass~\cite{Wang_CPC2021}.}
\end{center}
\end{table*}

In all six cases studied using PXCT~(Table~\ref{PXCT_history}), only the average lifetimes of proton-unbound states populated by EC were obtained. Individual proton-emitting states could not be fully resolved due to high level densities. Moreover, the applicability of this technique has not been explored in an astrophysical context. We have designed, built, and tested LIBRA to extend PXCT to measure most essential ingredients for calculating reaction rates with the Hauser-Feshbach statistical model Eq.~\eqref{eq:HauserFeshbach}~\cite{Rauscher_ADNDT2000}. LIBRA may also be able to identify individual resonances, providing spins and parities, excitation/resonance energies, lifetimes, and $p,\alpha,\gamma$ decay branching ratios for calculating narrow resonance reaction rates using Eqs.~\eqref{eq:ReactionRate} and~\eqref{eq:ResonanceStrength}.

EC/$\beta^+$ ratios are energy dependent, and in the center of the $^{59}$Cu$(p,\gamma)^{60}$Zn Gamow window, when a 1-MeV resonance is populated by allowed $^{60}$Ga $\beta$ transitions, $R_{\mathrm{EC}/\beta^+}=2.0\times10^{-3}$~\cite{Chen_RadiationReport}. The fractional probability of EC occurring in the $K$ shell is 89\%~\cite{Chen_RadiationReport}. The $K$-shell fluorescence yield for Zn is 47\%, with the remaining 53\% attributed to Auger electrons~\cite{Perkins_EADL1991}. Additionally, $K_{\alpha1}$ and $K_{\alpha2}$ X rays constitute 59\% and 30\%, respectively, of the total $K$ X-ray emission~\cite{Perkins_EADL1991}. Since their energies are generally indistinguishable in experiments, we collectively refer to these X rays as $K_{\alpha}$. Combining these decay probabilities, we estimate that for a 1-MeV resonance in $^{60}$Zn populated in $^{60}$Ga decay, approximately $7.4\times10^{-4}$ of such events will produce X rays suitable for lifetime analysis.

\section{Experimental Setup}
\subsection{Beam delivery}
For the future experiment case study under consideration, the Facility for Rare Isotope Beams (FRIB) linear accelerator~\cite{Wei_MPLA2022} will accelerate $^{70}$Ge to 249~MeV/nucleon. The reaction products from $^{70}$Ge impinging on a rotating carbon transmission target will be separated in flight by the Advanced Rare Isotope Separator~\cite{Portillo_NIMB2023}. A cocktail fast beam containing $^{60}$Ga and some nearby isotones will be slowed down in metal degraders with momentum compression and thermalized in gas stoppers filled with helium~\cite{Sumithrarachchi_NIMB2020,Lund_NIMB2020}. The thermalized $^{60}$Ga ions will be drifted by a combination of radio-frequency and direct-current fields towards a nozzle and exit into a radio-frequency quadrupole ion-guide system. The ions will be guided and accelerated to 30~keV before being delivered to the stopped beam area~\cite{Villari_NIMB2023}. The scientific user program involving stopped and reaccelerated beams started in 2023, and the beam intensities will continue increasing as the primary beam power gradually ramps up to 400~kW over the coming years. The FRIB beam rate calculator yields a stopped-beam rate of $^{60}$Ga of $3\times10^5$~pps. It should be noted that the calculator assumes optimal conditions and actual rates are likely to be lower than the calculated values~\cite{FRIB_BeamCalc}.

\begin{figure*}[htbp!]
\begin{center}
\includegraphics[width=16cm]{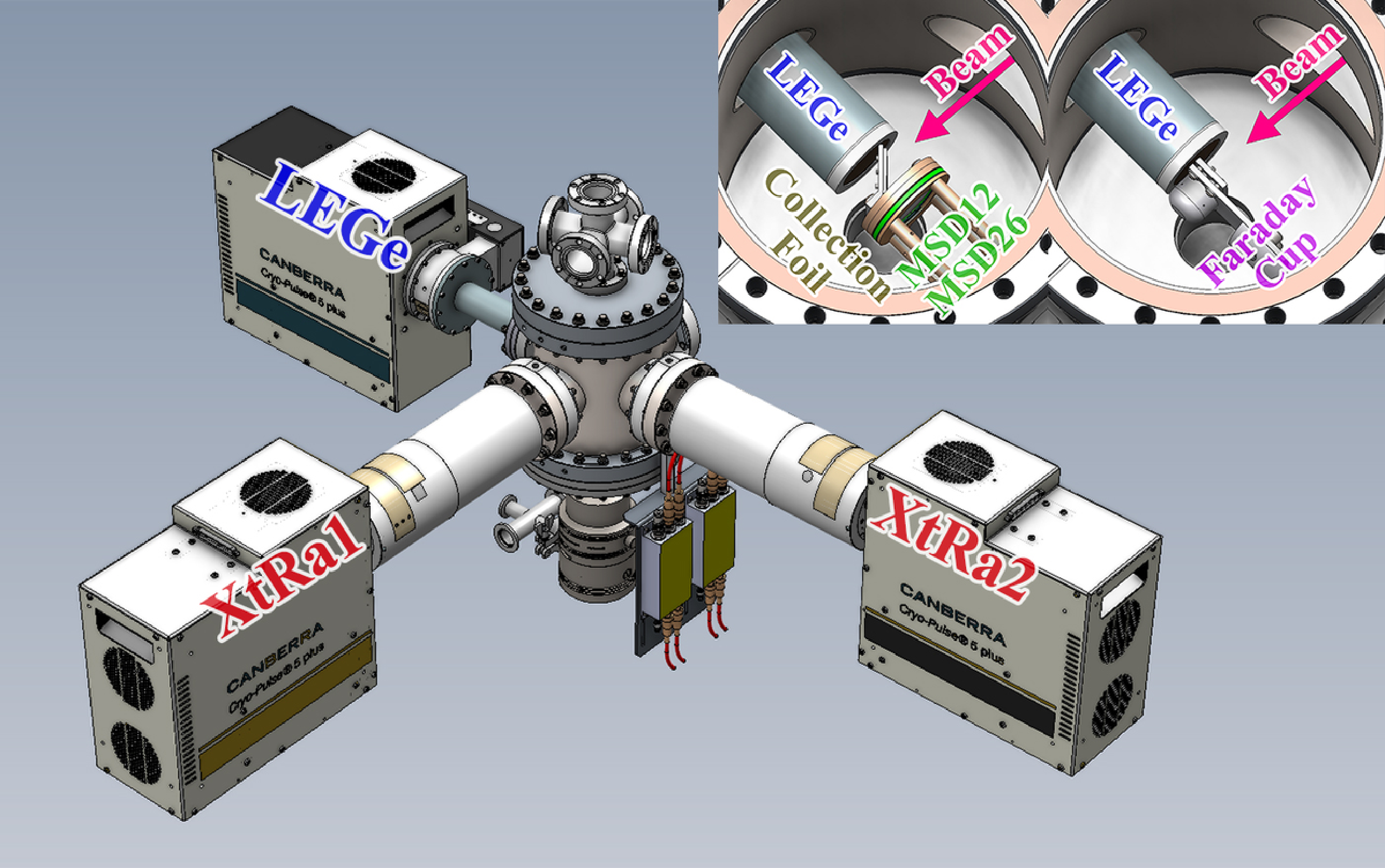}
\includegraphics[width=16cm]{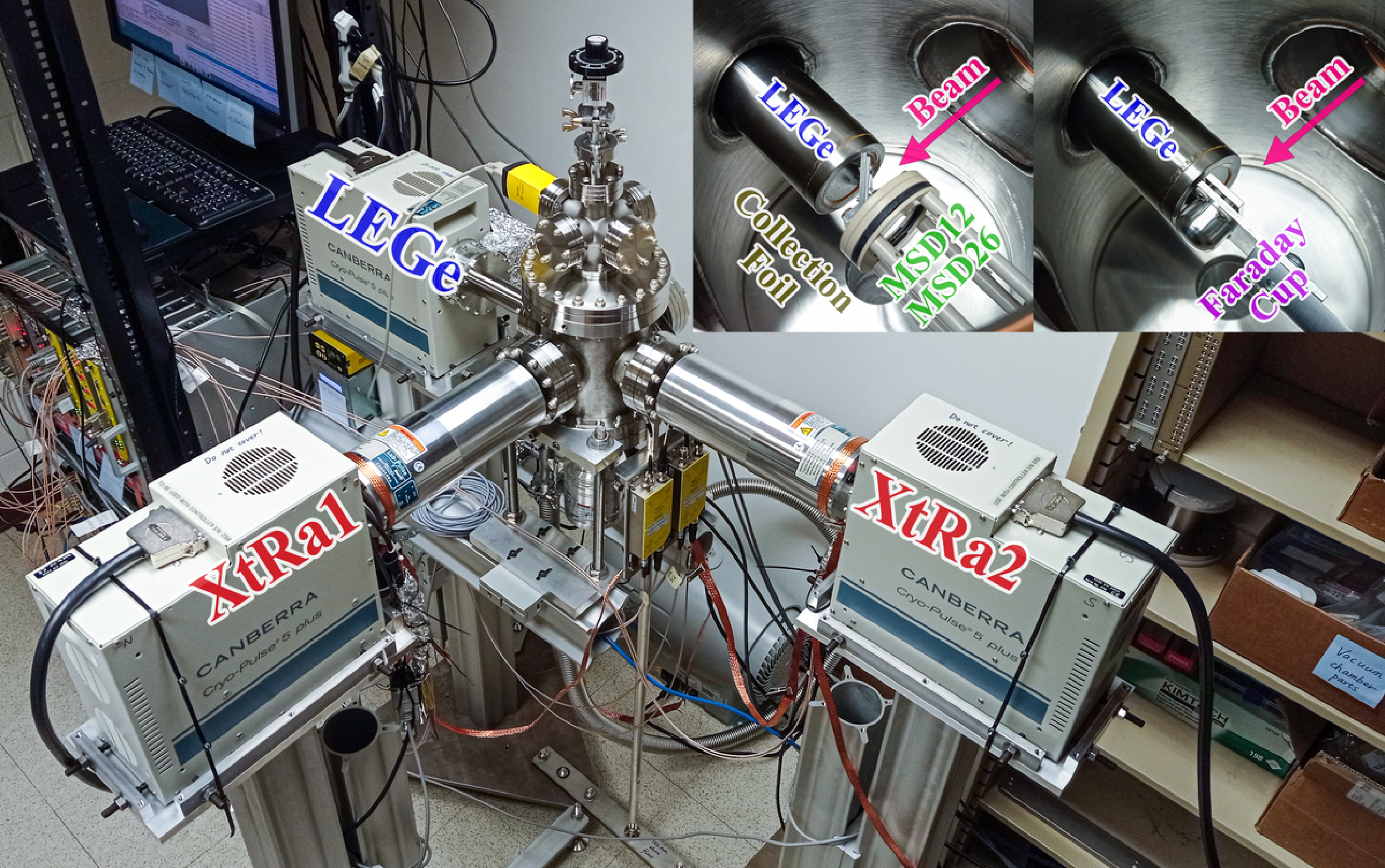}
\caption{\label{PXCT_setup} Mechanical design drawing and photograph of LIBRA. The insets highlight two configurations for the detectors inside the central chamber: a Faraday cup with a collimator for beam tuning or a collection foil and Si detectors for decay measurements.}
\end{center}
\end{figure*}

A mechanical design drawing and photograph of LIBRA are shown in Fig.~\ref{PXCT_setup}. Prior to the experiment, a stable beam around the $A=60$ region will be tuned into the Faraday cup at the center of the vacuum chamber. After maximizing the beam current, the chamber will be vented and the Faraday cup will be replaced by a thin aluminized Mylar foil tilted at a 45$^{\circ}$ angle with respect to the beam direction. The $^{60}$Ga beam will then be directed into the center of the foil using the previously established beam tune. A 30-keV $^{60}$Ga beam can be fully stopped by a Mylar foil as thin as 50~nm~\cite{Ziegler_NIMB2010}, in contrast to the 6.5~mm needed to stop a 130-MeV/nucleon $^{60}$Ga fast beam~\cite{E23035_GADGET}. Since fast beams penetrate deeply into materials and would block the emitted X rays and charged particles, we have chosen to use stopped beams instead of fast beams to effectively utilize PXCT.


\subsection{Detectors}
The detection system consists of a planar germanium detector for X-ray detection, two large-volume coaxial germanium detectors for $\gamma$-ray detection, and a silicon telescope for charged-particle detection via energy-loss and residual energy ($\mathrm{\Delta}E$-$E$).
 
For X-ray detection, we selected a disk-shaped Low Energy Germanium detector (LEGe), Mirion GL0510~\cite{MIRION_LEGe}. The LEGe detector comprises a Ge crystal with a diameter of 25.0~mm and a thickness of 10.5~mm. LEGe is housed in a flanged-style cryostat with a diameter of 38.1~mm and a 0.13-mm thick Be entrance window. The endcap is inserted into the vacuum chamber with its entrance window 11.0~mm from the center of the chamber. The Ge crystal is positioned 5.6~mm from the entrance window, subtending 10.1\% of the $4\pi$ solid angle with respect to the center of the chamber. LEGe is fabricated with a thin $p^+$ contact on the front and side, and a rear $n^+$ contact that covers less than the full area, resulting in lower capacitance than a similar-sized planar device. Since preamplifier noise is a function of detector capacitance, the low capacitance feature makes LEGe ideally suited for X-ray spectroscopy down to 3~keV.

For $\gamma$-ray detection, we selected two Extended Range Coaxial Germanium Detectors (XtRa), Mirion GX10020~\cite{MIRION_XtRa}. The active volume of XtRa1 has a diameter of 84.8~mm and a thickness of 65.2~mm, while XtRa2 has a diameter of 79.8~mm and a thickness of 80.0~mm. The Ge crystals are positioned 6.8 and 6.3~mm, respectively, from their 0.6-mm-thick carbon composite windows. The XtRa detectors feature a thin window contact on the front surface and an $n^+$ contact on the periphery, providing a good low-energy response.

All three Ge detectors are equipped with the Cryo-Pulse 5 Plus electrically refrigerated cryostat~\cite{MIRION_CP5,Willems_Mirion2015}. The detector housing is connected to a compact cold-head assembly containing a 5-watt pulse tube cooler. The assembly is powered by a bench-top controller, which contains the necessary logic to ensure the safe and reliable operation of the cryostat. During normal operations, the cold tip is maintained at the preset $-185~^{\circ}$C. Additionally, a control panel application provides remote control, monitoring, and logging of cryostat status.


For the $\mathrm{\Delta}E$-$E$ charged-particle telescope, we selected two single-sided, single-area circular Si detectors manufactured by Micron Semiconductor Ltd. The active volume of MSD12 is 12~$\mathrm{\mu}$m in thickness and 12~mm in diameter~\cite{MICRON_MSD12}, and MSD26 is 1000~$\mathrm{\mu}$m thick and 26~mm in diameter~\cite{MICRON_MSD26}. The junction side of both MSDs features a 50-nm thick boron-doped silicon dead layer and a 30-$\mu$m wide peripheral metal band for wire bonding, leaving the majority of the active area without metal coverage. The Ohmic side of MSD12 has a thicker dead layer of 300~nm with no metal coverage. The Ohmic side of MSD26 has little impact on charged-particle signals, and thus, we opt for the standard 500-nm thick dead layer and 300-nm thick aluminum coverage. Both silicon chips are assembled onto an FR4 printed circuit board. MSD26 is positioned 15.7~mm from the center of the chamber and covers 11.5\% of the 4$\pi$ solid angle. MSD12 is 11.2~mm from the center and defines the solid angle coverage of the $\mathrm{\Delta}E$-$E$ telescope at 5.9\% of 4$\pi$.


\subsection{Electronics}
All three Ge detectors are equipped with Intelligent Preamplifiers (iPA)~\cite{MIRION_iPA}, which incorporate a low-noise field-effect transistor (FET) input circuit optimized for the ultra-high source impedance of Ge detectors. The first stage of the iPA functions as an integrator, providing an output voltage proportional to the accumulated charge. The second stage of the iPA acts as an output buffer and offers four selectable gain settings. The output signal is split into two channels with termination impedances of 93 and 50~$\Omega$, respectively. The iPA memory stores detector leakage currents, temperatures, and preamplifier operating voltages. A control panel application allows for remote monitoring and logging of these parameters. Each iPA is equipped with two 100-$\Omega$ Pt resistance temperature detectors thermally connected to the crystal holder (PRTD1) and the cold tip (PRTD2), respectively~\cite{Eberth_EPJA2023}. The PRTD1 reading represents the temperature of the Ge crystal when they are in thermal equilibrium. If either PRTD exceeds its preset threshold, it can trigger the high-voltage inhibit via the iPA. This mechanism operates independently of the inhibit function via the controller, providing enhanced protection for the detector.

Two ORTEC~660 Dual Bias Supply modules~\cite{ORTEC660} are used to provide bias voltages to the three Ge detectors. We apply a negative bias to the $p^+$ contacts of LEGe and a positive bias to the $n^+$ contacts of XtRa. LEGe becomes fully depleted at $-$600~V and is recommended to be operated at $-$1100~V. XtRa1 and XtRa2 become fully depleted at bias voltages of $+$4000~V and $+$2200~V, respectively, and both operate at $+$4500~V. The bias shutdown mode of the ORTEC~660 is set to transistor-transistor logic (TTL) to be compatible with the iPA high-voltage inhibit mode. The typical leakage currents of the two XtRa detectors are below 20~pA and below 100~pA for LEGe. The tail pulses from iPAs exhibit rise times of $\approx$150~ns (LEGe) and $\approx$250~ns~(XtRa), with a 50-$\mathrm{\mu}$s decay constant.

A Mesytec MHV 4-channel bias supply module with remote control features provides the bias voltages to the two MSD Si detectors. We apply a negative bias to the $p^+$ contacts of both MSD detectors through MPR-1 charge-sensitive preamplifiers~\cite{MPR1}, and the $n^+$ contacts are grounded. MSD12 has a depletion voltage of $-$1.5~V and is operated at $-$3.0~V, and MSD26 has a $-$90~V depletion voltage and is operated at $-$130~V. MHV offers a ramp speed as low as 5~V/s to protect the circuits of preamplifiers~\cite{MHV4}. MSD12 exhibits a leakage current of approximately 1~nA, while MSD26 exhibits a leakage current of approximately 60~nA. The energy and timing outputs of the MPR-1 are both terminated with 50~$\Omega$ impedance. The tail pulses from MPR-1 exhibit rise times of $\approx$400~ns (MSD12) and $\approx$70 ns~(MSD26), with a 120~$\mathrm{\mu}$s decay constant. All preamplifiers are powered by two Mesytec MNV-4 NIM power distribution and control modules~\cite{MNV4}. 

\subsection{Data acquisition}
All preamplifier signals are transmitted through double-shielded RG316 coaxial cables of equal length and then digitized by a 16-bit, 250~MHz Pixie-16 module manufactured by XIA LLC~\cite{XIA_Pixie}. The input impedance of each channel in Pixie-16 is switchable between 50~$\Omega$ and 1~k$\Omega$. The Digital Data Acquisition System (DDAS)~\cite{Starosta_NIMA2009,Prokop_NIMA2014} is used for recording and processing data. Trapezoidal filtering algorithms are implemented in both the slow filter for pulse amplitude measurement and the fast filter for leading-edge triggering. The DDAS filter parameters are optimized based on Refs.~\cite{Prokop_NIMA2014,XIA_Pixie_Manual,Wu_NIMA2020,Weisshaar_NIMA2017}.

The system operates in an internally triggered mode: recording data on a channel-by-channel basis whenever the trigger filter crosses the user-defined threshold. Data from all channels are ordered in time and subsequently assembled into events in software based on a user-defined event window length. Each event is timestamped using a Constant Fraction Discriminator (CFD) algorithm based on the trigger filter response. The event timestamp is counted with 125~MHz clock ticks, i.e., 8~ns intervals.

The pulse amplitude is extracted from the energy filter amplitude at approximately rise time plus gap time after triggering. If a second trigger arrives within the rise time plus gap time window, both events will be flagged as pileup. In the tests conducted in the next section, pileup rejection is turned off in DDAS. As a result, the timestamp of the first event is preserved while the second is discarded for all the tests conducted in the next section. The amplitude of the second event is partially added to that of the first event, with the addition diminishing as the time interval between them increases. The count rate capacity of the detection system is primarily determined by the energy filter parameters selected for each detector.

\begin{figure}[htbp!]
\begin{center}
\includegraphics[width=8.6cm]{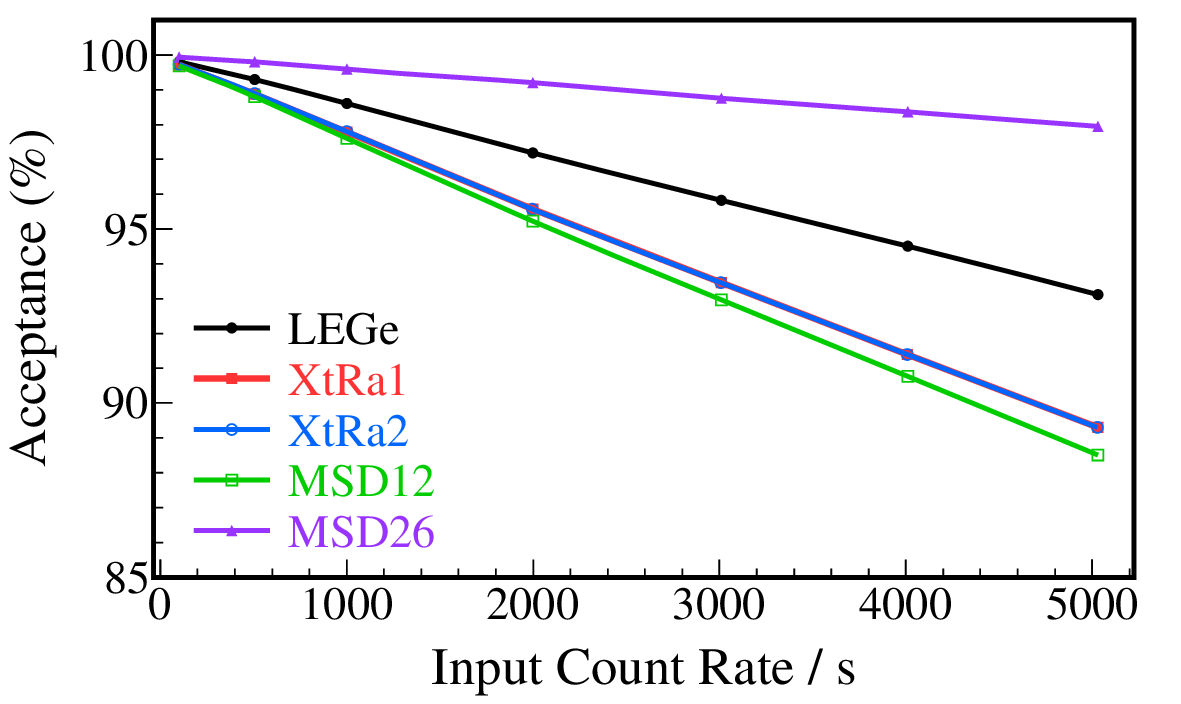}
\caption{\label{DDAS_Dead_Time} Count rate performance of each detector.}
\end{center}
\end{figure}

To assess DDAS live time under our parameter setting, a DB-2 Random Pulser~\cite{BNC_DB2Pulser} is used to generate pulses with time intervals following a Poisson distribution. The recorded count rates are shown in Fig.~\ref{DDAS_Dead_Time}, and are consistent with the pileup rates defined by the energy filter settings~\cite{Starosta_NIMA2009}. Furthermore, DDAS can provide real-time spectra for identifying characteristic charged particles and $\gamma$ rays from decay, aiding in the online identification of radioactive beams.


\section{Performance Tests}
We performed offline tests on LIBRA using the radioactive sources listed in Table~\ref{Sources}. An event-build window of $\pm1~\mathrm{\mu}$s was set, and the count rate of each detector remained below 1500~events per second throughout all conducted tests, except for a LEGe test with the $^{152}$Eu source.

\begin{table}
\caption{\label{Sources} Characteristics of the radioactive sources used in the LIBRA detector tests. Columns 2$-$7 present the source nuclides, main decay modes, actual activities ($A$), relative uncertainties of the activities ($\sigma$$A$), active diameters ($D$), and half-lives ($T_{1/2}$), respectively.}
\begin{center}
\renewcommand{\arraystretch}{1.2}  
\begin{ruledtabular}
\begin{tabular}{ccccccc}
No. & Nuclide & Decay & $A$ (Bq) & $\sigma$$A$ (\%) & $D$ (mm) & $T_{1/2}$ (yr)\footnotemark[1]\\
\hline
1 & $^{55}$Fe & EC & $1.11\times10^4$ & $-$\footnotemark[2] & 9.5 & 2.756 \\
2 & $^{60}$Co & $\beta^-$ & $3.73\times10^4$ & 3 & 1 & 5.271 \\
3 & $^{137}$Cs & $\beta^-$ & $3.00\times10^3$ & 3 & 3 & 30.007 \\
4 & $^{148}$Gd & $\alpha$ & $2.86\times10^4$ & $-$\footnotemark[2] & 5 & 86.9 \\
5 & $^{152}$Eu & EC/$\beta^-$ & $3.10\times10^4$ & 1.4 & 3 & 13.517 \\
6 & $^{241}$Am & $\alpha$ & $3.44\times10^3$ & 2.7 & 3 & 432.6 \\
\end{tabular}
\end{ruledtabular}
\footnotetext[1]{Adopted from Ref.~\cite{NWC2023}, omitting uncertainties.}
\footnotetext[2]{Unknown source activity uncertainties; not used for efficiency calibration.}
\end{center}
\end{table}

\subsection{X-ray measurements}
\begin{figure*}[htbp!]
\begin{center}
\includegraphics[width=17cm]{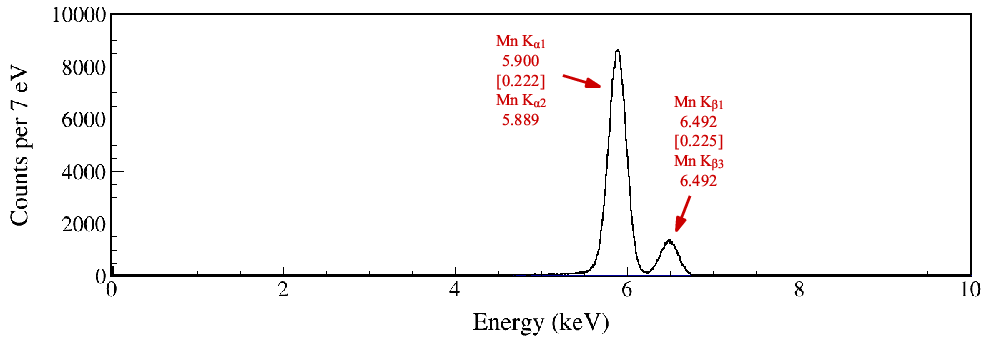}
\includegraphics[width=17cm]{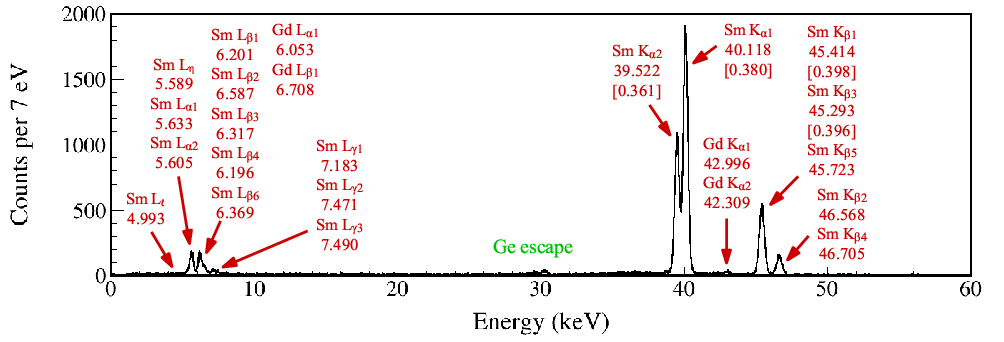}
\includegraphics[width=17cm]{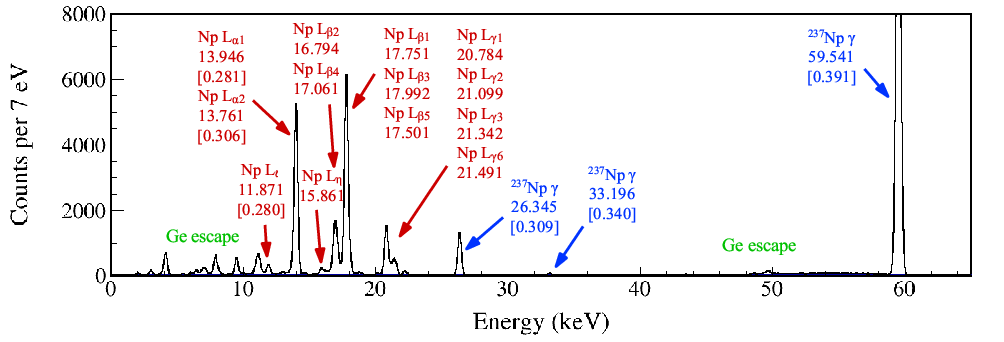}
\caption{\label{LEGe_X_spec} X-ray and/or $\gamma$-ray spectra measured by the LEGe detector using the $^{55}$Fe (top), $^{152}$Eu (middle), and $^{241}$Am (bottom) sources. X-ray energy values are adopted from Ref.~\cite{Newville_XrayDB2023} rounded to the nearest 0.001~keV. $\gamma$-ray energy values are adopted from Ref.~\cite{Basunia_NDS2006_A237} rounded to the nearest 0.001~keV. FWHM values used to characterize the energy resolution of LEGe are indicated within brackets.}
\end{center}
\end{figure*}

We evaluated the performance of LEGe using the $^{55}$Fe, $^{152}$Eu, and $^{241}$Am sources, as shown in Fig.~\ref{LEGe_X_spec}. $^{55}$Fe EC decays to $^{55}$Mn ground state, and the subsequent filling of atomic shell vacancies results in X rays characteristic of Mn. Similarly, Sm X rays mainly result from $^{152}$Eu EC. $^{152}$Eu decay populates $^{152}$Sm/$^{152}$Gd excited states, which can deexcite via internal conversion (IC), followed by filling of atomic shell vacancies and the emission of X rays characteristic of Sm/Gd. This explains why the observed Gd X rays are much weaker compared to Sm X rays that have two production mechanisms: EC and IC. For $^{241}$Am, $\alpha$ decay populates $^{237}$Np excited states, where IC serves as the primary mechanism leading to Np X rays. A trace amount of X rays may also arise from inner-shell ionization and excitation caused by perturbations in the electron cloud during nuclear decays~\cite{Isozumi_NIMA1989,Freedman_ARNPS1974,Crasemann_1975}. The 0.13-mm-thick Be entrance window is sufficient to block electrons below 125~keV~\cite{Kantele_1995}, rendering the LEGe detector insensitive to Auger electrons.

The overall energy resolution achieved by LEGe is characterized by fitting X-ray or $\gamma$-ray lines with an exponentially modified Gaussian (EMG) function~\cite{Sun_PRC2021} to account for incomplete charge collection at 5.90~keV (Mn $K_{\alpha1}$), 6.49~keV (Mn $K_{\beta1}$), 11.89~keV (Np $L_{\ell}$), 13.76~keV (Np $L_{\alpha2}$), 13.95~keV (Np $L_{\alpha1}$), 26.34~keV ($^{237}$Np $\gamma$), 33.20~keV ($^{237}$Np $\gamma$), 39.52~keV (Sm $K_{\alpha2}$), 40.12~keV (Sm $K_{\alpha1}$), 45.29~keV (Sm $K_{\beta3}$), 45.41~keV (Sm $K_{\beta1}$), and 59.54~keV ($^{237}$Np $\gamma$). The manufacturer specifies the full width at half maximum (FWHM) values for LEGe as 0.218~keV at 5.9~keV ($^{55}$Fe) and 0.514~keV at 122~keV ($^{57}$Co), respectively. Figure~\ref{LEGe_X_spec} demonstrates that the observed energy resolution aligns with these specifications. We then interpolated the FWHM values at the energies of interest, 8.05~keV (Cu $K_{\alpha1}$) and 8.64~keV (Zn $K_{\alpha1}$), to be 0.238(8) and 0.241(7)~keV, respectively, demonstrating sufficient resolution to distinguish between the key X rays of Zn and Cu in our case study.

For photons below 100~keV interacting with Ge, the photoelectric effect is predominant, i.e., the photon is absorbed, and a photoelectron is ejected by the Ge atom. When the resulting atomic shell vacancy is filled, X rays characteristic of Ge may be created. A full-energy peak is still observed if these X rays are reabsorbed near the original interaction site. However, if the photoelectric interaction occurs near the surface of Ge, the X rays are more likely to escape, which results in peaks usually at 9.89 and 10.98~keV below the photopeaks, known as the Ge escape peaks (Fig.~\ref{LEGe_X_spec}). These energy differences correspond to the characteristic $K_{\alpha}$ and $K_{\beta}$ X-ray energies for Ge, respectively~\cite{Newville_XrayDB2023}.

\begin{figure}[htbp!]
\begin{center}
\includegraphics[width=8.5cm]{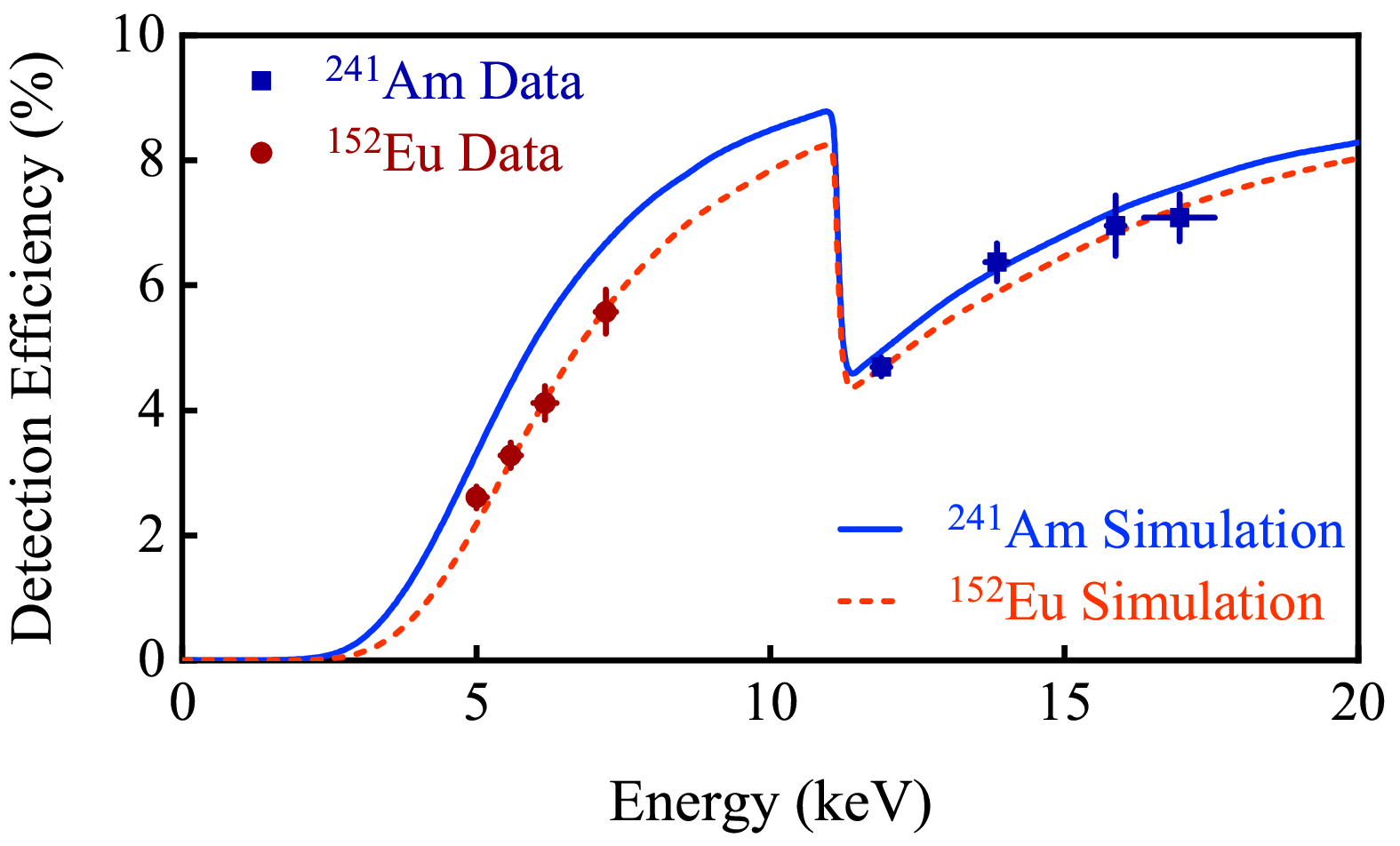}
\caption{\label{LEGe_XEfficiency}Absolute X-ray photopeak detection efficiency of the LEGe detector obtained using the Sm $L_\ell$, $L_\eta + L_\alpha$, $L_\beta$, and $L_\gamma$ X rays from the $^{152}$Eu source and Np $L_\ell$, $L_\alpha$, $L_\eta$, and $L_\beta$ X rays from the $^{241}$Am source, each placed at the center of the chamber. The red dashed and blue solid curves represent the \textsc{geant}4 simulated efficiencies according to the $^{152}$Eu and $^{241}$Am source configurations, respectively. The error bars along the x-axis also reflect the energy span for the multiple X rays within each group.}
\end{center}
\end{figure}

We evaluated the detection efficiency of LEGe using the X rays from the $^{152}$Eu source placed at the center of the chamber tilted at a 45$^{\circ}$ angle with respect to LEGe. $^{152}$Eu emits Sm $L$ X rays at 5.0~keV ($L_\ell$), 5.6~keV ($L_\eta$, $L_\alpha$), 6.2~keV ($L_\beta$), and 7.2~keV ($L_\gamma$). The Gd $L$ X rays are approximately half a keV higher but with two orders of magnitude lower intensities. We adopted the total $L$ X-ray emission probability from Ref.~\cite{Be_TOR2004} and deduced the absolute intensities for each of the 4 groups of X rays based on the relative emission probabilities reported by Ref.~\cite{Mehta_NIMA1986}. The corresponding efficiencies are indicated by the four low-energy data points in Fig.~\ref{LEGe_XEfficiency}. We also measured the X rays from the $^{241}$Am source placed at the center of the chamber. $^{241}$Am emits Np $L$ X rays at 11.9~keV ($L_\ell$), 13.9~keV ($L_\alpha$), 15.9~keV ($L_\eta$), and 17.0~keV ($L_\beta$)~\cite{Be_TOR2010}. The corresponding efficiencies are indicated by the four high-energy data points in Fig.~\ref{LEGe_XEfficiency}.

We simulated the X-ray detection efficiencies using \textsc{geant}4~\cite{Agostinelli_NIMA2003,Allison_NIMA2016}. The simulation incorporates the geometric configuration of the setup and the LEGe detector response, which was characterized by fitting the measured X-ray lineshapes in Fig.~\ref{LEGe_X_spec} with the EMG function. Monoenergetic X rays are emitted isotropically from the source position and interact with the surrounding materials. The simulation outputs an energy spectrum, from which we obtain the detection efficiency by dividing the counts in the X-ray peak by the number of emitted X rays. This process was repeated at different energies to generate the efficiency curves shown in Fig.~\ref{LEGe_XEfficiency}.

For photon energies just above the $K$-shell binding energy of Ge, 11.1030(20)~keV~\cite{Newville_XrayDB2023}, the incident photon is strongly absorbed without deep penetration beyond the detector surface. The subsequent characteristic $K$ X rays of 9.7$-$11.1~keV tend to escape. However, for photons just below the Ge $K$-shell binding energy, $K$-shell absorption is no longer possible, and $L$-shell interactions dominate. In this case, incident photons tend to penetrate somewhat deeper, and the chance of escape of the fluorescent Ge $L$ X rays of 1.0$-$1.4~keV is significantly lower. This phenomenon abruptly changes the full-energy detection efficiency of X rays near the $K$-shell absorption edge~\cite{Knoll_2010}. The $^{241}$Am source used for this test is an open source, while the $^{152}$Eu source is encapsulated between two 60-$\mathrm{\mu}$m-thick Mylar tapes. The Mylar layer attenuates low-energy X-rays, but its impact diminishes for X rays above 10~keV. Additionally, the LEGe count rate was $\approx$3000~pps during the $^{152}$Eu test but only $\approx$200~pps during the $^{241}$Am test, resulting in different DDAS live time (Fig.~\ref{DDAS_Dead_Time}). Therefore, the $^{152}$Eu efficiency curve represents a lower limit, while the $^{241}$Am efficiency curve represents an ideal setting. The $^{60}$Ga experimental condition is expected to fall between these two scenarios, and we estimate the X-ray efficiencies at 8.0 and 8.6~keV to be 6.5$-$7.4\% and 7.0$-$7.8\%, respectively.

\subsection{$\gamma$-ray measurements}
\begin{figure*}[htbp!]
\begin{center}
\includegraphics[width=17cm]{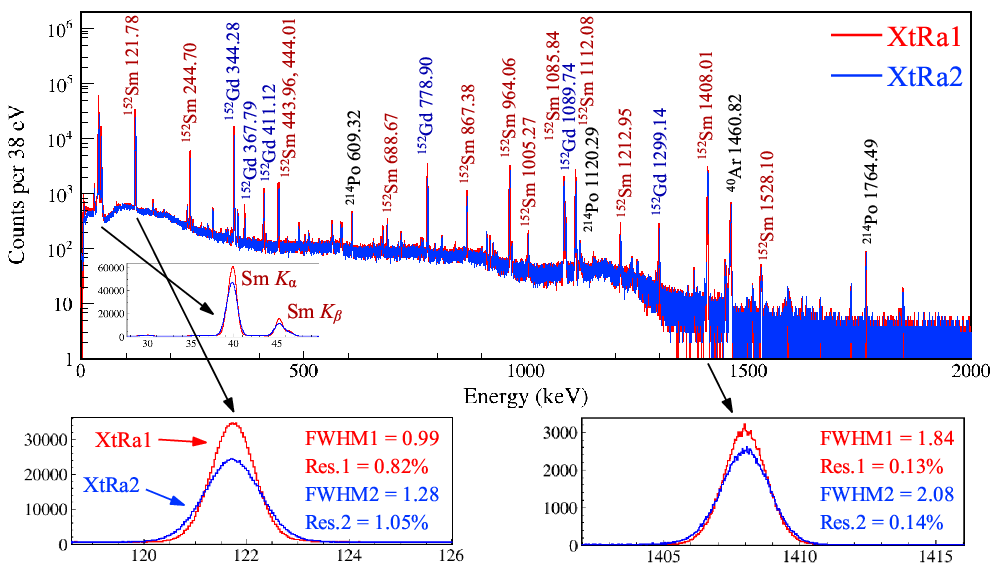}
\includegraphics[width=17cm]{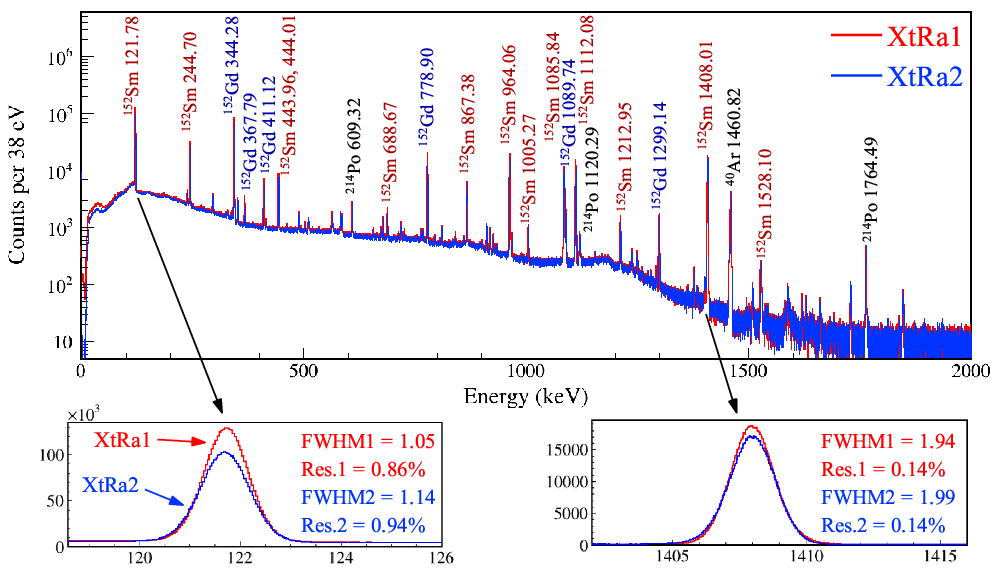}
\caption{\label{XtRa_Gamma_spec} $\gamma$-ray spectra measured by XtRa1 (red) and XtRa2 (blue) using the $^{152}$Eu source. Upper panel: the $^{152}$Eu source is placed in the middle of the two XtRa facing each other. Lower panel: the $^{152}$Eu source is placed at the center of the vacuum chamber, with the two XtRa detectors positioned according to the Fig.~\ref{PXCT_setup} configuration. All the $\gamma$-ray energy values are adopted from Ref.~\cite{Martin_NDS2013_A152} rounded to the nearest 0.01~keV. The insets demonstrate the detector responses at 122 and 1408~keV.}
\end{center}
\end{figure*}

Figure~\ref{XtRa_Gamma_spec} shows the $\gamma$-ray spectra measured by XtRa1 and XtRa2 using the $^{152}$Eu source. We first placed the source at the midpoint between the two XtRa detectors that were facing each other, with a distance of 28~cm between them. Both XtRa detectors exhibit good low-energy response to the $^{152}$Sm X rays at 40~keV. We then placed the source at the center of the vacuum chamber to determine the absolute $\gamma$-ray detection efficiencies. The two XtRa detectors were placed as close as possible to the two flanges (Fig.~\ref{PXCT_setup}), with their entrance windows about 12~mm from the flange surface. The XtRa1 Ge crystal has a slightly larger diameter than that of XtRa2. Both Ge crystals are 158.5~mm from the target center, covering 1.70\% and 1.51\% of the 4$\pi$ solid angle, respectively. Both XtRa detectors record an average of $\approx$300 room background $\gamma$ rays per second in our laboratory test environment. The manufacturer specifies the FWHM values for XtRa1 and XtRa2 as 0.998 and 1.065~keV at 122~keV ($^{57}$Co), and 1.879 and 1.926~keV at 1332~keV ($^{60}$Co), respectively. The insets of Fig.~\ref{XtRa_Gamma_spec} demonstrate that the observed energy resolution using the $^{152}$Eu source aligns with these specifications. The absence of X-ray peaks in the second test (lower panel of Fig.~\ref{XtRa_Gamma_spec}) is due to the 3.175-mm-thick stainless steel flanges of the chamber effectively blocking the X rays.

We also measured the $\gamma$-ray detection efficiencies using the $^{60}$Co and $^{137}$Cs sources placed at the center of the chamber. MSD12 was not in place during these tests due to its fragility. MSD26 and the Si detector holders attenuated the $\gamma$ rays from the source to XtRa2 but had little effect on XtRa1. Based on an exponential function that contains a polynomial of degree $i$ with the natural logarithm of the energy $E$, $\varepsilon(E)=\exp\left[\sum_{i=0}^{6}p_i\ln(E)^i\right]$~\cite{Glassman_PRC2019}, fit on all the data points, we obtain the photopeak efficiencies of 0.334(3)\% and 0.286(3)\% at 1~MeV, respectively, for XtRa1 and XtRa2. The error bars on the data points reflect the uncertainty of the $\gamma$-ray yields and the source activities, with an additional 2.5\% uncertainty to account for the true coincidence summing effect~\cite{Gilmore_2008,Palazzo_PRL2023}, which was estimated based on the observed 1173-1332-keV $\gamma$ cascade from $^{60}$Co.

\begin{figure}[htbp!]
\begin{center}
\includegraphics[width=8.6cm]{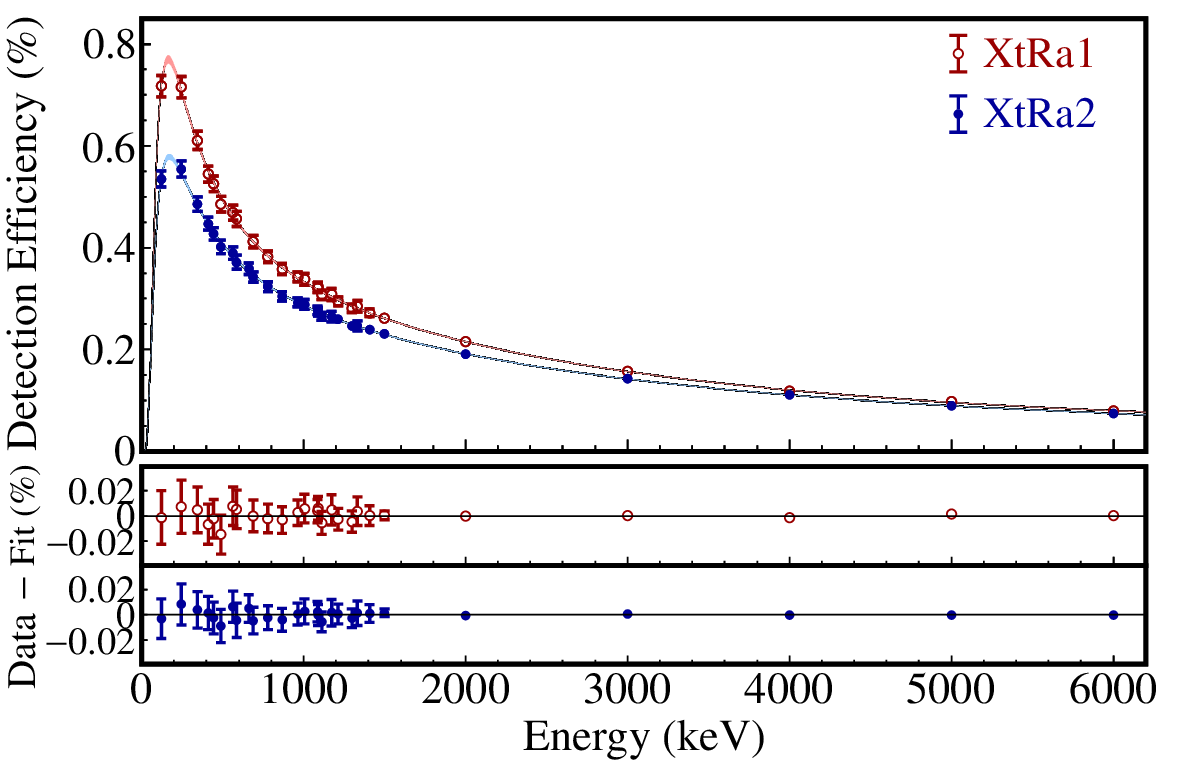}
\caption{\label{XtRa_Efficiency}Absolute $\gamma$-ray photopeak detection efficiency of the two XtRa detectors obtained using the $^{152}$Eu, $^{137}$Cs, and $^{60}$Co sources placed at the center of the chamber. The $^{137}$Cs data point at 662~keV is only applicable to XtRa2 due to the source placement. The six data points above 1408~keV are \textsc{geant}4 simulated efficiencies scaled by a factor to match the low-energy source data. The efficiency curves are generated by fitting all measured and simulated data points.}
\end{center}
\end{figure}

We used \textsc{geant}4 simulation~\cite{Agostinelli_NIMA2003,Allison_NIMA2016} to extend the $\gamma$-ray detection efficiency curve to high energies (Fig.~\ref{XtRa_Efficiency}). The simulation takes into account the geometry of the setup and the detector response characterized by fitting the measured $\gamma$-ray lineshapes with the EMG function. Monoenergetic $\gamma$ rays were emitted isotropically according to the source distribution and interacted with the surrounding materials. The photopeak efficiency was extracted from the output spectrum. We then fit the ratio of the simulated efficiency to the measured efficiency in the range 0.5$-$1.5~MeV and obtained energy-independent ratios of 0.875(10) and 0.837(10) for XtRa1 and XtRa2, respectively, which serve as the normalization factors to match the simulation with the experimental data. One of the factors that reduces the measured efficiency is the data acquisition event loss, which is estimated to be 3.3\%, 0.7\%, and 2.1\% based on the count rates during the $^{60}$Co, $^{137}$Cs, and $^{152}$Eu tests, respectively~(Fig.~\ref{DDAS_Dead_Time}).

\subsection{Charged-particle measurements}
\begin{figure}[htbp!]
\begin{center}
\includegraphics[width=8.6cm]{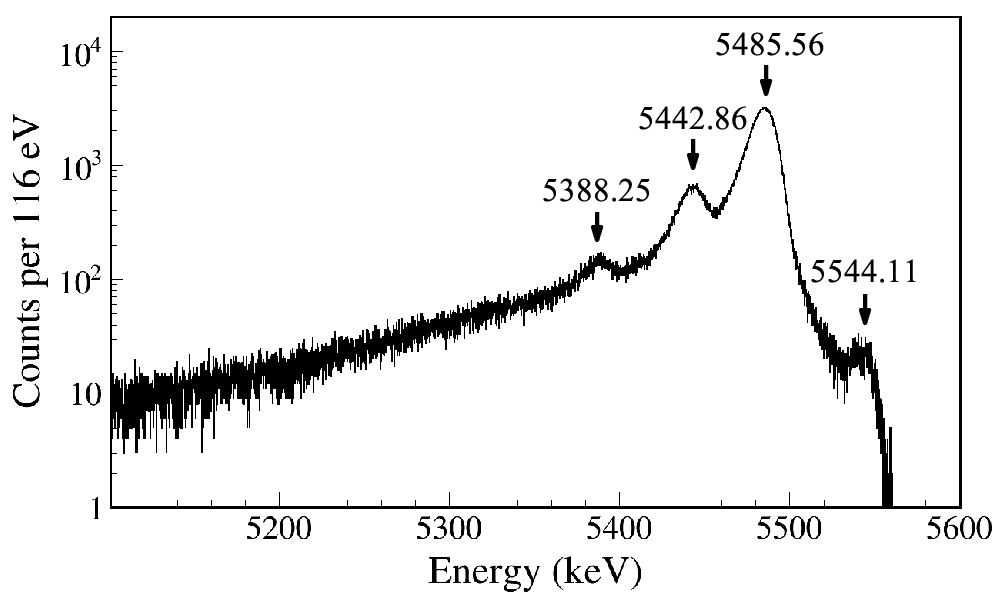}
\caption{\label{MSD26_Alpha_spec} $\alpha$ spectrum measured by MSD26 using the $^{241}$Am source. The $\alpha$ energy values are adopted from Ref.~\cite{Be_TOR2010} rounded to the nearest 0.01~keV. The FWHM value at 5485.56~keV is 17.0~keV, corresponding to an energy resolution of 0.31\%.}
\end{center}
\end{figure}

\begin{figure}[htbp!]
\begin{center}
\includegraphics[width=8.5cm]{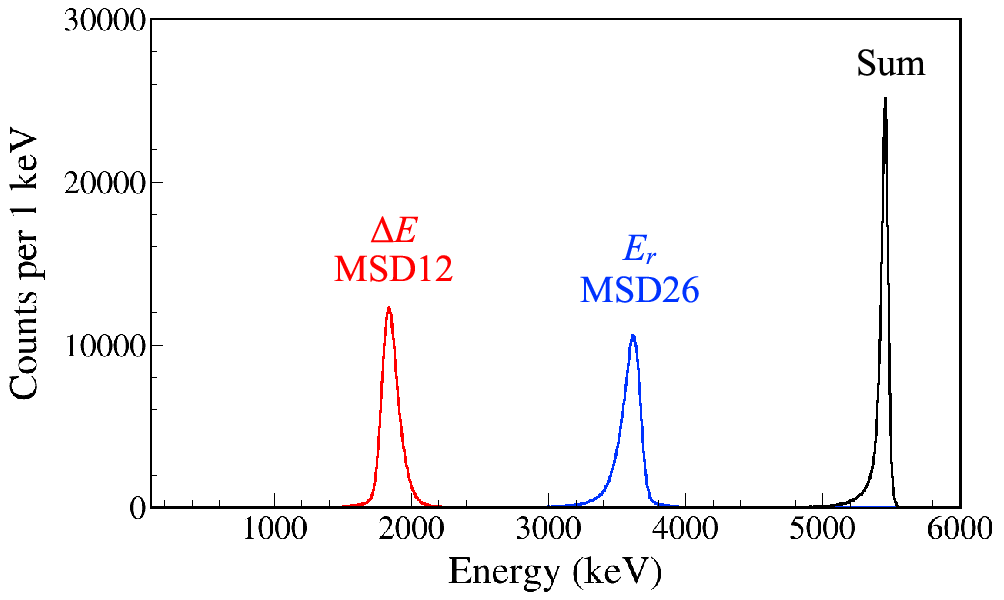}
\includegraphics[width=8.5cm]{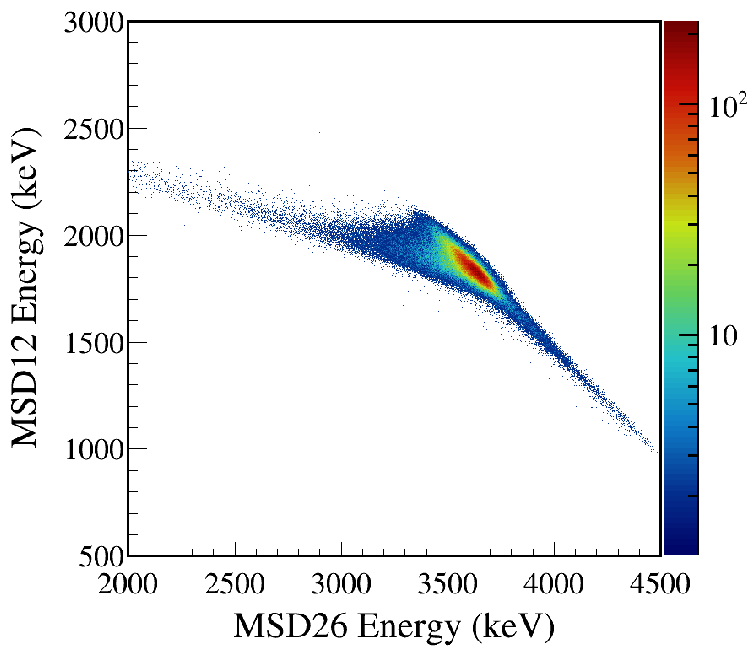}
\caption{\label{MSD_DE_E_PID} Upper panel: $^{241}$Am $\alpha$-energy spectra measured by MSD12 (energy loss) and MSD26 (residual energy). The FWHM value of the sum peak is 52.1~keV, corresponding to an energy resolution of 0.95\%. Lower panel: $\Delta E$-$E$ 2D plot.}
\end{center}
\end{figure}

Figure~\ref{MSD26_Alpha_spec} shows the $\alpha$ spectrum measured by MSD26 alone using the $^{241}$Am source, with a 2-mm-diameter aperture installed in front. An EMG fit of the main peak at 5485.56~keV yields a FWHM value of 17.0~keV, corresponding to an energy resolution of 0.31\%. For comparison, the manufacturer specifies the FWHM values for MSD26 as 26.7 and 35.4~keV, respectively, with different bias settings. MSD12 alone is too thin to stop $\alpha$ particles above 3~MeV, and we demonstrate the $\Delta E$-$E$ $\alpha$ spectra measured by the telescope formed by MSD12 and MSD26 in Fig.~\ref{MSD_DE_E_PID}. An EMG fit of the energy-sum peak yields a FWHM value of 52.1~keV, corresponding to an energy resolution of 0.95\%. This resolution is better than that of the Si telescopes used in the previous $^{60}$Ga experiment~\cite{Mazzocchi_EPJA2001}, where FWHM values for $^{148}$Gd ($E_\alpha=3182.68$~keV~\cite{Akovali_NDS1998}) were reported to be 100 and 90~keV for the $\Delta E$ and 50 and 60~keV for the $E$ detectors.

MSD26 was calibrated using the $^{148}$Gd and $^{241}$Am sources. We then measured the residual energy of $^{241}$Am $\alpha$ particles in MSD26 with MSD12 installed in front of it. This allowed us to accurately determine the effective thickness of MSD12 to be 11.65(8)~$\mathrm{\mu}$m after subtracting the 0.35 $\mathrm{\mu}$m dead layer thickness~\cite{Ziegler_NIMB2010}. The total thickness of MSD12 is in agreement with the nominal value of 12~$\mathrm{\mu}$m specified in the Micron datasheet~\cite{MICRON_MSD12}.

\subsection{Electron measurements}
\begin{figure}[htbp!]
\begin{center}
\includegraphics[width=8.6cm]{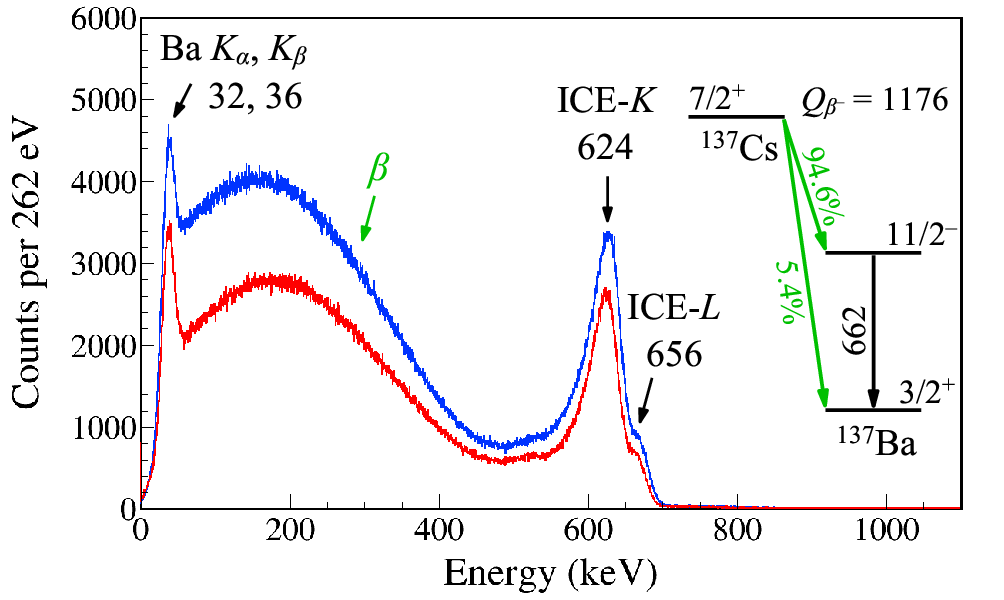}
\caption{\label{MSD26_Electron_spec}Electron spectra measured by MSD26 using the $^{137}$Cs source. The spectrum with lower statistics (red) was obtained with MSD12 installed between the source and MSD26. The spectrum with higher statistics (blue) was acquired over an equal time period with MSD12 removed. Electrons from $^{137}$Cs $\beta^-$ decay form the continuum. ICE-$K$ and ICE-$L$ denote the internal conversion electrons ejected from Ba $K$ and $L$ atomic shells, respectively. The low-energy peak is mainly from Ba $K_\alpha$ X rays at 32~keV and $K_\beta$ X rays at 36~keV. All energy values are adopted from Ref.~\cite{Mougeot_Metrologia2025} rounded to the nearest keV. A simplified $^{137}$Cs decay scheme shows the main decay branches.}
\end{center}
\end{figure}

Figure~\ref{MSD26_Electron_spec} shows the electron spectra measured by MSD26 using the $^{137}$Cs source placed at the center of the chamber facing MSD26. The source is deposited on a 64.4-$\mu$m-thick aluminized Mylar disk and covered with a 6.3-$\mu$m-thick Kapton window. The spectrum exhibits a continuum of electrons from $^{137}$Cs $\beta^-$ decay, along with distinct electron peaks from IC. The main $\beta^-$ decay branch has an endpoint energy of 514~keV and the IC peaks are characterized by the energy differences between the 662-keV $^{137}$Ba isomeric transition and the Ba atomic shell binding energies. Using the counts in the IC peaks measured by MSD26 alone, the total IC electron emission intensity of 9.56(14)\% per $^{137}$Cs decay~\cite{Mougeot_Metrologia2025}, and the source activity~(Table~\ref{Sources}), we estimate the detection efficiency of MSD26 for $^{137}$Ba IC electrons to be 9.0(3)\%.

\subsection{Coincidence measurements}
\begin{figure}[htbp!]
\begin{center}
\includegraphics[width=8.5cm]{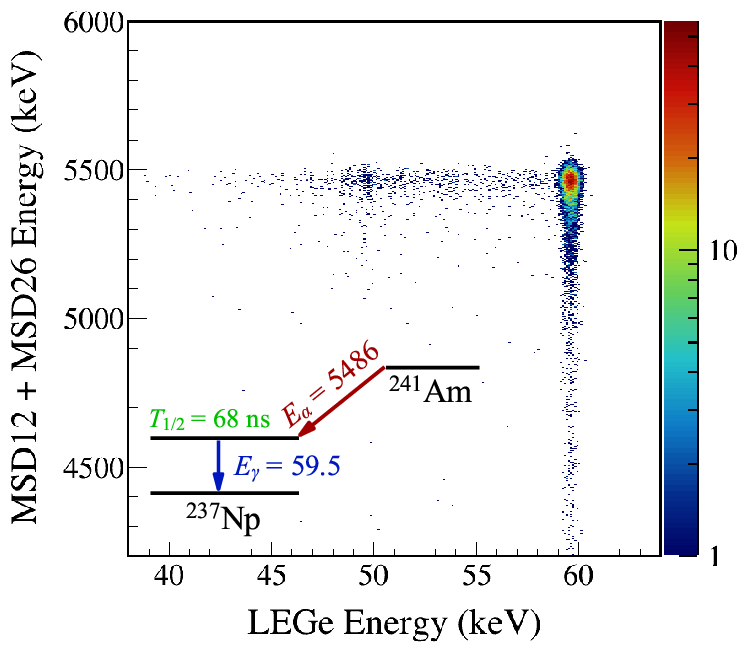}
\caption{\label{alphagamma_coin_spec_241Am} Coincidence spectrum between the MSD detector telescope and LEGe obtained using the $^{241}$Am source placed at the center of the chamber. A simplified $^{241}$Am decay scheme shows the dominant $\alpha$-$\gamma$ sequence.}
\end{center}
\end{figure}

Figure~\ref{alphagamma_coin_spec_241Am} shows the $\alpha$-$\gamma$ coincidence spectrum between the MSD telescope and LEGe with the $^{241}$Am source placed at the center of the chamber. The source faces the MSD, and its 127-$\mu$m-thick Pt substrate attenuates most of the low-energy photons emitted towards LEGe, leaving mainly the 59.5-keV $^{237}$Np $\gamma$ ray and its escape peaks observable.

\begin{figure*}[htbp!]
\begin{center}
\includegraphics[width=17cm]{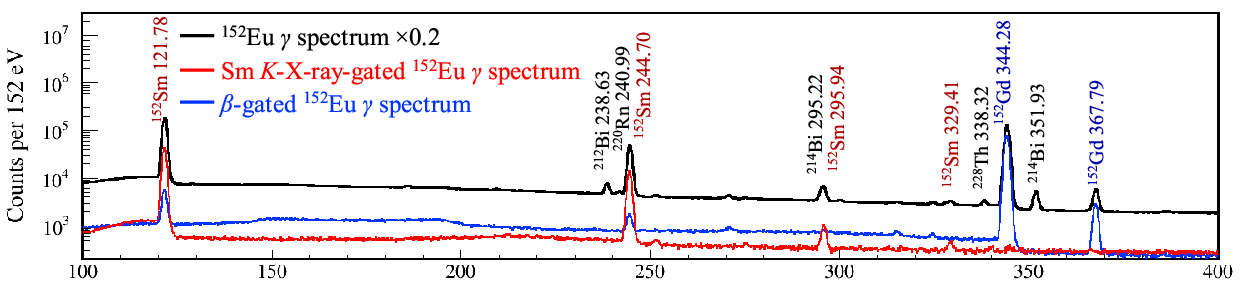}
\includegraphics[width=17cm]{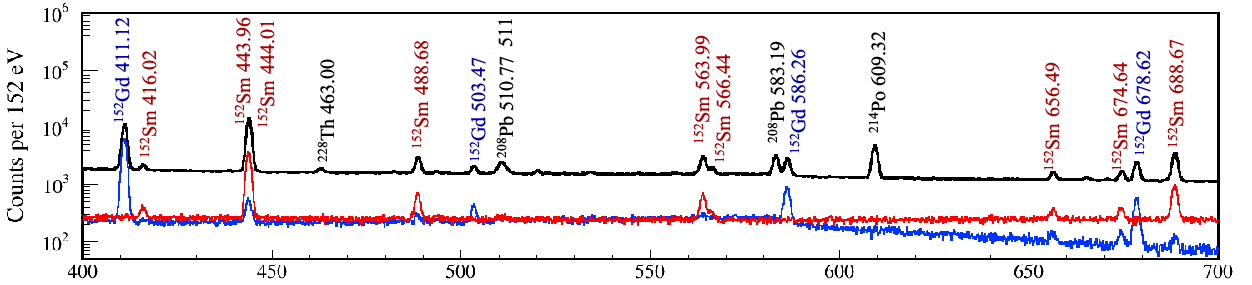}
\includegraphics[width=17cm]{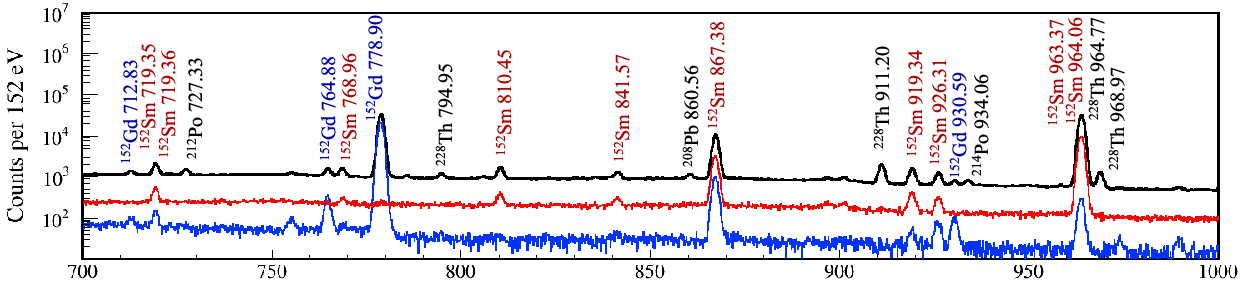}
\includegraphics[width=17cm]{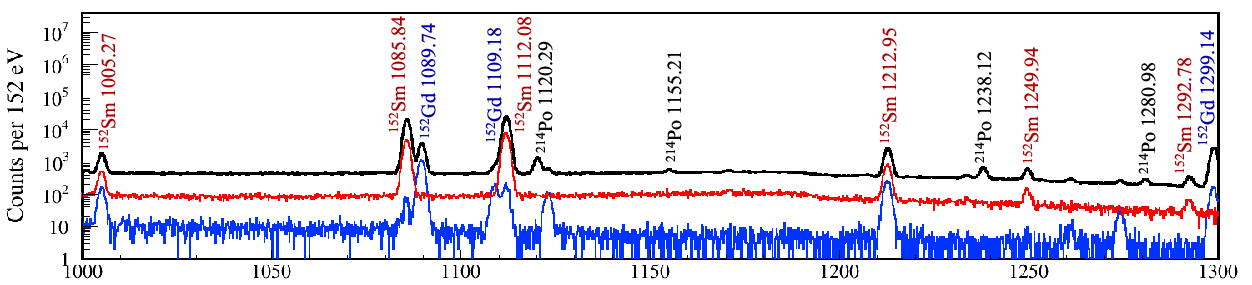}
\includegraphics[width=17cm]{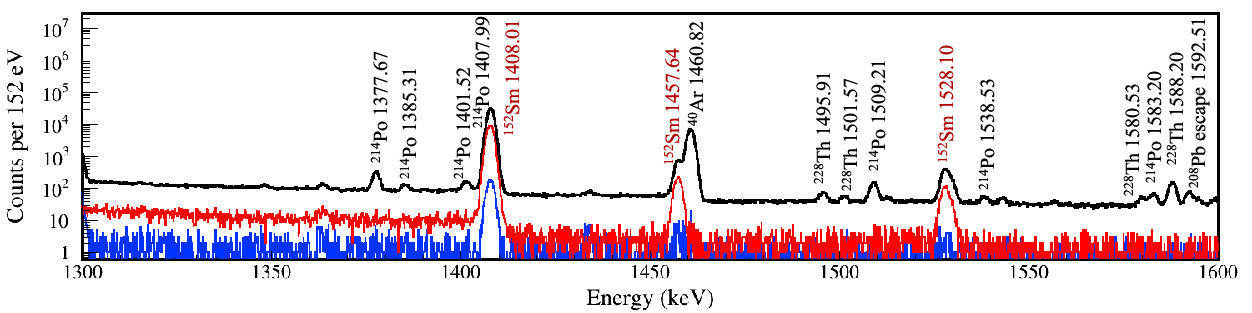}
\caption{\label{xgamma_coin_spec_152Eu} Black spectrum with the highest statistics represents the raw $\gamma$-ray spectrum measured by XtRa1 using the $^{152}$Eu source placed at the center of the chamber. The red spectrum with intermediate statistics represents the XtRa1 $\gamma$-ray spectrum gated by the Sm $K_\alpha$ and $K_\beta$ X rays measured by LEGe. The blue spectrum with the lowest statistics represents the XtRa1 $\gamma$-ray spectrum gated by the electrons measured by MSD26. The raw (black) spectrum is scaled down by a factor of 5 for better comparison.}
\end{center}
\end{figure*}

We placed the $^{152}$Eu source at the center of the chamber. Figure~\ref{xgamma_coin_spec_152Eu} shows the XtRa1 $\gamma$ spectra gated by the Sm $K$ X rays measured by LEGe and gated by the electrons measured by MSD26. By applying the characteristic X-ray coincidence condition, both the room background $\gamma$ rays and the $^{152}$Gd $\gamma$ rays are substantially suppressed. Conversely, the electron coincidence condition suppresses the room background and the $^{152}$Sm $\gamma$ rays. Quantitatively, at 1 MeV, the background levels per 152-eV bin decrease from approximately 625 counts in the raw spectrum to 21 and 3 counts in the X-ray-gated and electron-gated spectra, respectively. Having the ability to detect electrons and positrons will help clean up the spectrum in radioactive beam measurements, thereby facilitating the identification of $\gamma$ ray origins.

\subsection{Timing performance}
The timing performance of the electronics was first tested using a Canberra Model 1407P Pulse Pair Generator~\cite{Canberra1407P}. The dual pulses were separately fed into two Pixie-16 channels. The FWHM resolution of the time-difference distribution is estimated to be 0.46~ns. Then, the primary pulse was split and fed to each test input of preamplifiers, and the resulting FWHM timing resolutions are 37.4~ns (MSD12), 4.4~ns (MSD26), 1.2~ns (XtRa1), and 1.8~ns (XtRa2).

\begin{figure}[htbp!]
\begin{center}
\includegraphics[width=8.5cm]{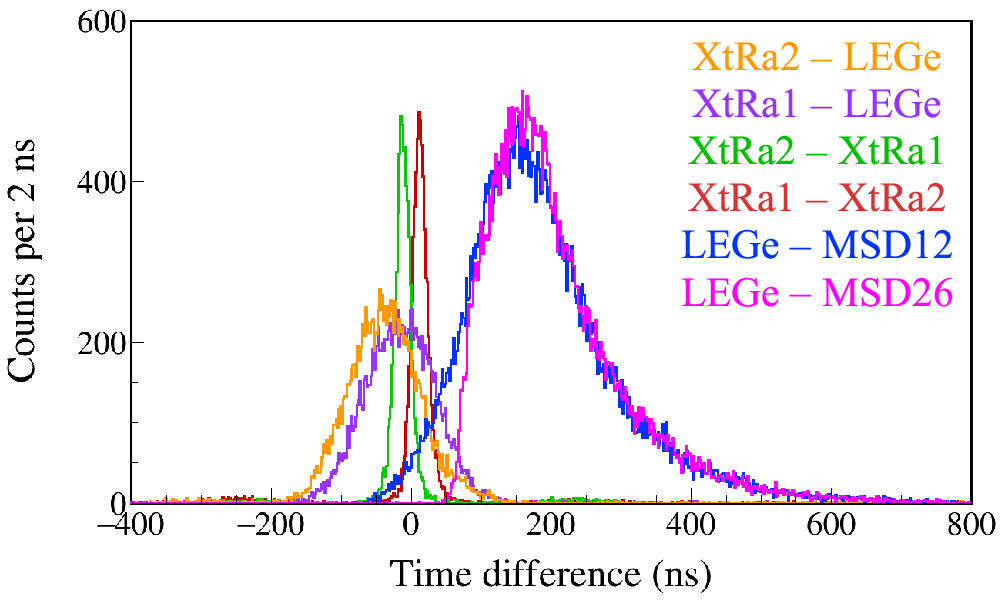}
\caption{\label{Timing_spec} Coincidence time spectra between each detector pair.  From left to right: the six time peaks correspond to three decay sequences: the $^{152}$Eu 40$-$46-keV and 1408-keV X-$\gamma$ coincidences measured by XtRa-LEGe, the $^{60}$Co 1173-keV and 1332-keV $\gamma$-$\gamma$ coincidences measured by XtRa-XtRa, and the $^{241}$Am 5486-keV and 59.5-keV $\alpha$-$\gamma$ coincidences measured by LEGe-MSD. In each decay sequence, the timestamp of the prior event is subtracted from the timestamp of the subsequent event.}
\end{center}
\end{figure}

The timing performance of the detectors was studied using each of the $^{60}$Co, $^{152}$Eu, $^{241}$Am sources placed at the center of the chamber. $^{60}$Co provides $\gamma$-$\gamma$ coincidences to test the two XtRa detectors, $^{152}$Eu provides X-$\gamma$ coincidences to test LEGe and XtRa, and $^{241}$Am provides $\alpha$-$\gamma$ coincidences to test MSD and LEGe. Figure~\ref{Timing_spec} shows the time difference distributions between each coincidence. Based on these measurements, an event-build window of a few hundred ns can be defined to capture all prompt coincidences and some chance continuum for background subtraction in offline analysis. The asymmetric tail in both $\alpha$-$\gamma$ time difference distributions is attributed to the relatively long-lived 59.5-keV excited state of $^{237}$Np.

\begin{figure}[htbp!]
\begin{center}
\includegraphics[width=8.5cm]{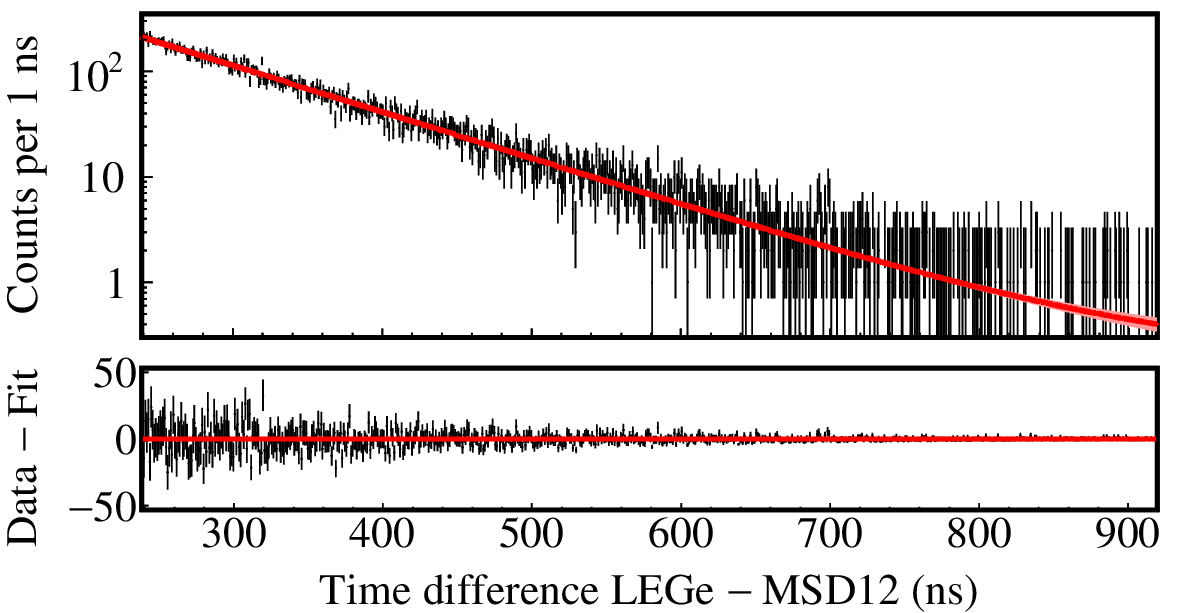}
\includegraphics[width=8.5cm]{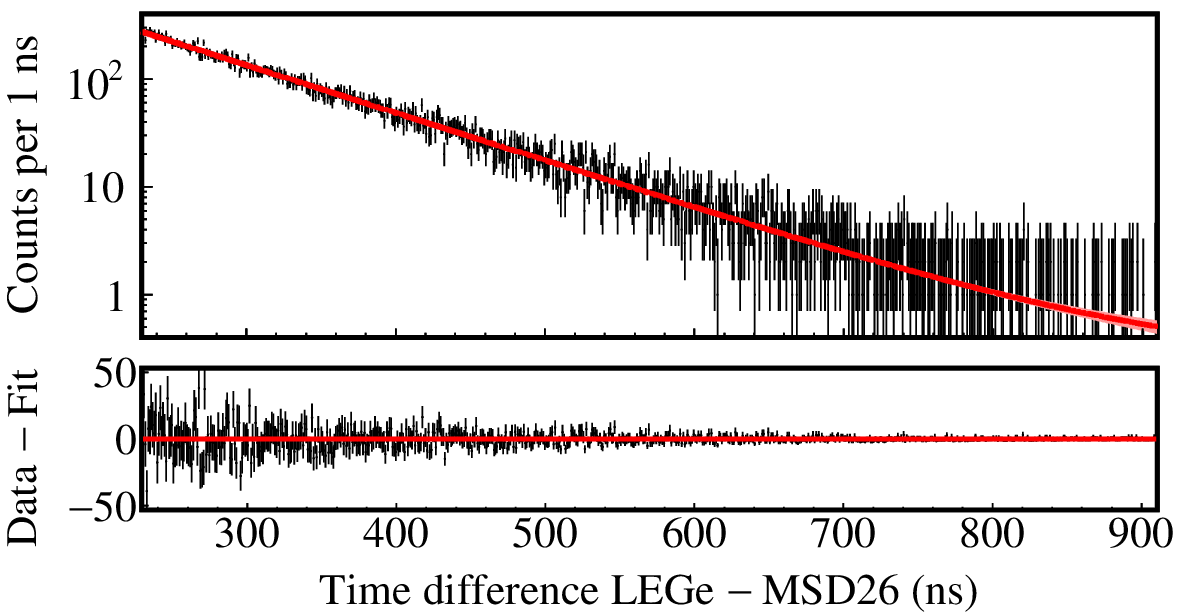}
\caption{\label{Timing_spec_241Am_alpha5486gamma59} Time differences between the 59.5-keV $^{237}$Np $\gamma$-ray signals in LEGe and the 5486-keV $^{241}$Am $\alpha$ signals in the MSD silicon detector telescope. By fitting the LEGe-MSD12 time difference (upper), we obtain the half-life of the second excited state of $^{237}$Np to be $T_{1/2}=68.1(6)$~ns with $p$ value = 0.34 and $\chi_\nu^2=1.02$, by dividing the $\chi^2$ value by the number of degrees of freedom. From the LEGe-MSD26 time difference (lower), we obtain $T_{1/2}=67.9(5)$~ns, $p$ value = 0.88, and $\chi_\nu^2=0.94$.}
\end{center}
\end{figure}

Figure~\ref{Timing_spec_241Am_alpha5486gamma59} shows the $\alpha$-$\gamma$ time difference distribution constructed by the start timestamps from 5486-keV $\alpha$ measured by the two MSDs and the stop timestamps from the 59.5-keV $\gamma$ ray deexciting the 59.5-keV state in $^{237}$Np measured by LEGe. By fitting the time spectra with a function

\begin{equation}
  f(t;N,T_{1/2},B) = \frac{N \ln(2)}{T_{1/2}} \exp\left[-\frac{t \ln(2)}{T_{1/2}}\right] + B
\end{equation}

composed of the total number of decays ($N$), the exponential decay half-life ($T_{1/2}$), and a constant background ($B$), we obtained the half-life of the 59.5-keV excited state in $^{237}$Np to be $T_{1/2}=68.1(6)$~ns (MSD12) and $67.9(5)$~ns (MSD26), respectively. The results obtained from both Si detectors are consistent with recent precision measurements of 67.86(9)~ns~\cite{Takacs_ARI2021}, 67.60(25)~ns~\cite{Dutsov_ARI2021}, and 67.60(20)~ns~\cite{Santos_NPA2023}.

In summary, the detectors selected for our setup all feature minimal or zero dead layers in front of their active regions. The mechanical assembly is designed to be compact and employs thin materials in the transmission path to mitigate attenuation, thereby maximizing detection efficiencies. Furthermore, the design also allows for the flexible combination of individual detectors for various experimental purposes. The two XtRa detectors have been coupled with a silicon cube~\cite{Jensen_Thesis2024,Eder_Thesis2024} and with a Time Projection Chamber~\cite{Mahajan_PRC2024}. We also have the option to engineer the integration of LEGe and the central chamber with larger detector arrays, such as the DEcay Germanium Array initiator (DEGAi) and ultimately DEGA of the FRIB Decay Station~\cite{FDS_WP2020}, to achieve a higher $\gamma$-ray detection efficiency.

\section{Calculations}

\begin{table*}[htbp!]
\caption{\label{Key_resonances}Properties of potentially important $^{59}$Cu$(p,\gamma)^{60}$Zn resonances predicted by the shell model. The values in the first through tenth columns represent the spin and parity ($J^\pi$), excitation energy ($E_x$), resonance energy ($E_r$), partial decay widths ($\Gamma_\gamma$, $\Gamma_p$, $\Gamma_\alpha$), lifetime ($\tau$), log~$ft$ value and $\beta$-feeding intensity ($I_\beta$) for $^{60}$Ga decay, and ratio of $\mathrm{EC}/\beta^+$ feeding~\cite{Chen_RadiationReport}.}
\begin{center}
\renewcommand{\arraystretch}{1.2}  
\begin{ruledtabular}
\begin{tabular}{cccccccccc}
$J^\pi$ & $E_x$~(keV) & $E_r$~(keV) & $\Gamma_\gamma$~(eV) & $\Gamma_p$~(eV) & $\Gamma_\alpha$~(eV)\footnotemark[1] & $\tau$~(fs) & log~$ft$ & $I_\beta$~(\%) & $R_{\mathrm{EC}/\beta^+}$ \\
\hline
$2^+$ & 5501 & 396 & $3.8\times10^{-2}$ & $7.4\times10^{-10}$ & $2.9\times10^{-7}$ & 17.3 & 5.463 & 0.314 & $1.6\times10^{-3}$ \\
$1^+$ & 5566 & 461 & $6.4\times10^{-2}$ & $1.5\times10^{-7}$ & 0 & 10.3 & 4.708 & 1.713 & $1.6\times10^{-3}$ \\
$2^+$ & 5645 & 540 & $1.9\times10^{-1}$ & $2.1\times10^{-6}$ & $1.1\times10^{-6}$ & 3.5 & 6.146 & 0.060 & $1.7\times10^{-3}$ \\
$2^+$ & 5989 & 884 & $3.3\times10^{-2}$ & $4.7\times10^{-3}$ & $1.6\times10^{-5}$ & 17.5 & 5.367 & 0.287 & $1.9\times10^{-3}$ \\
$2^+$ & 6072 & 967 & $2.5\times10^{-1}$ & $5.7\times10^{-2}$ & $2.9\times10^{-5}$ & 2.1 & 5.536 & 0.184 & $2.0\times10^{-3}$ \\
$1^+$ & 6305 & 1200 & $2.0\times10^{-1}$ & $2.1\times10^{-1}$ & $1.3\times10^{-27}$ & 1.6 & 7.035 & 0.005 & $2.2\times10^{-3}$ \\
\end{tabular}
\end{ruledtabular}
\footnotetext[1]{From the statistical model calculation.}
\end{center}
\end{table*}

\begin{figure}[htbp!]
\begin{center}
\includegraphics[width=8.5cm]{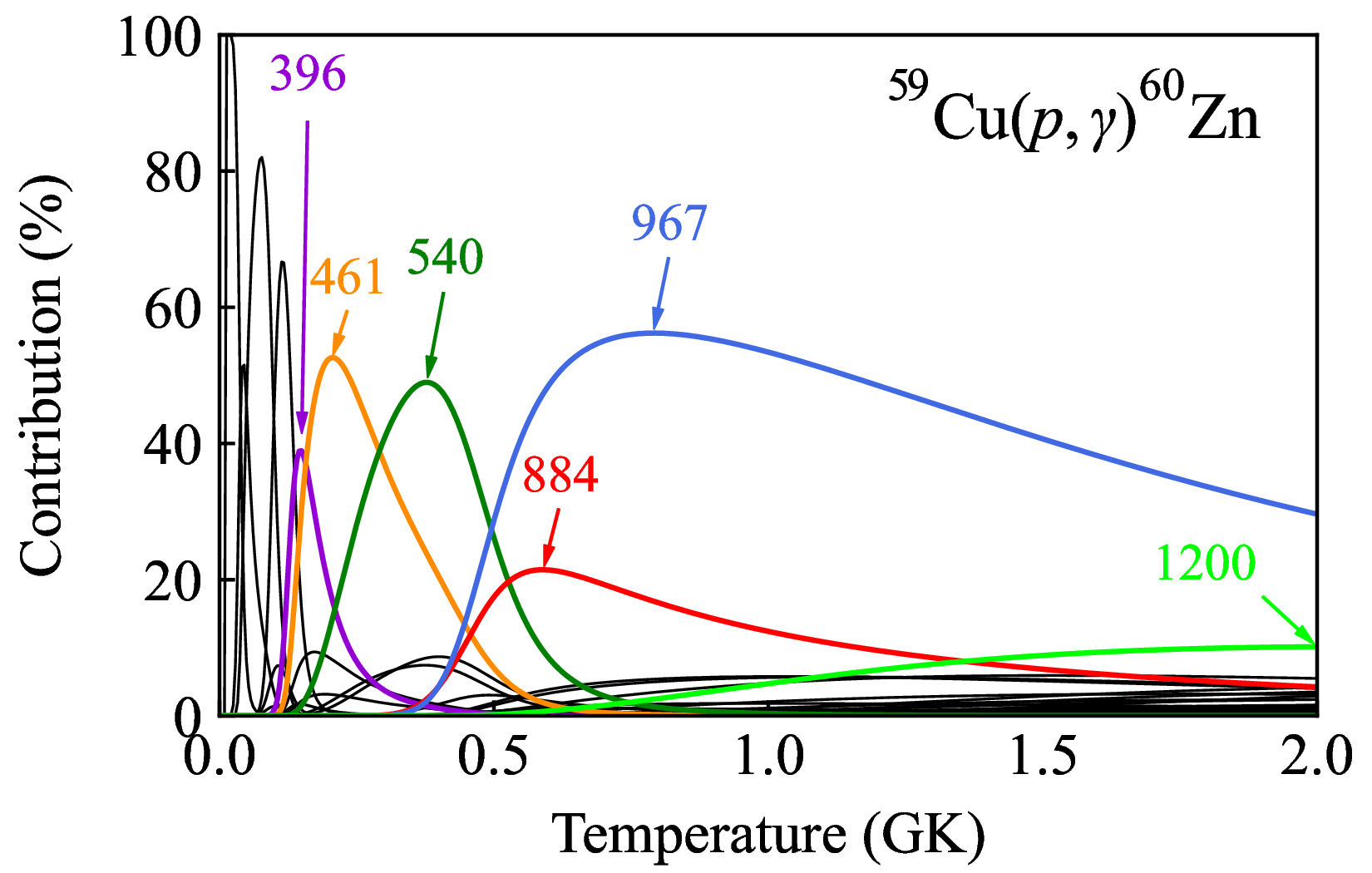}
\includegraphics[width=8.5cm]{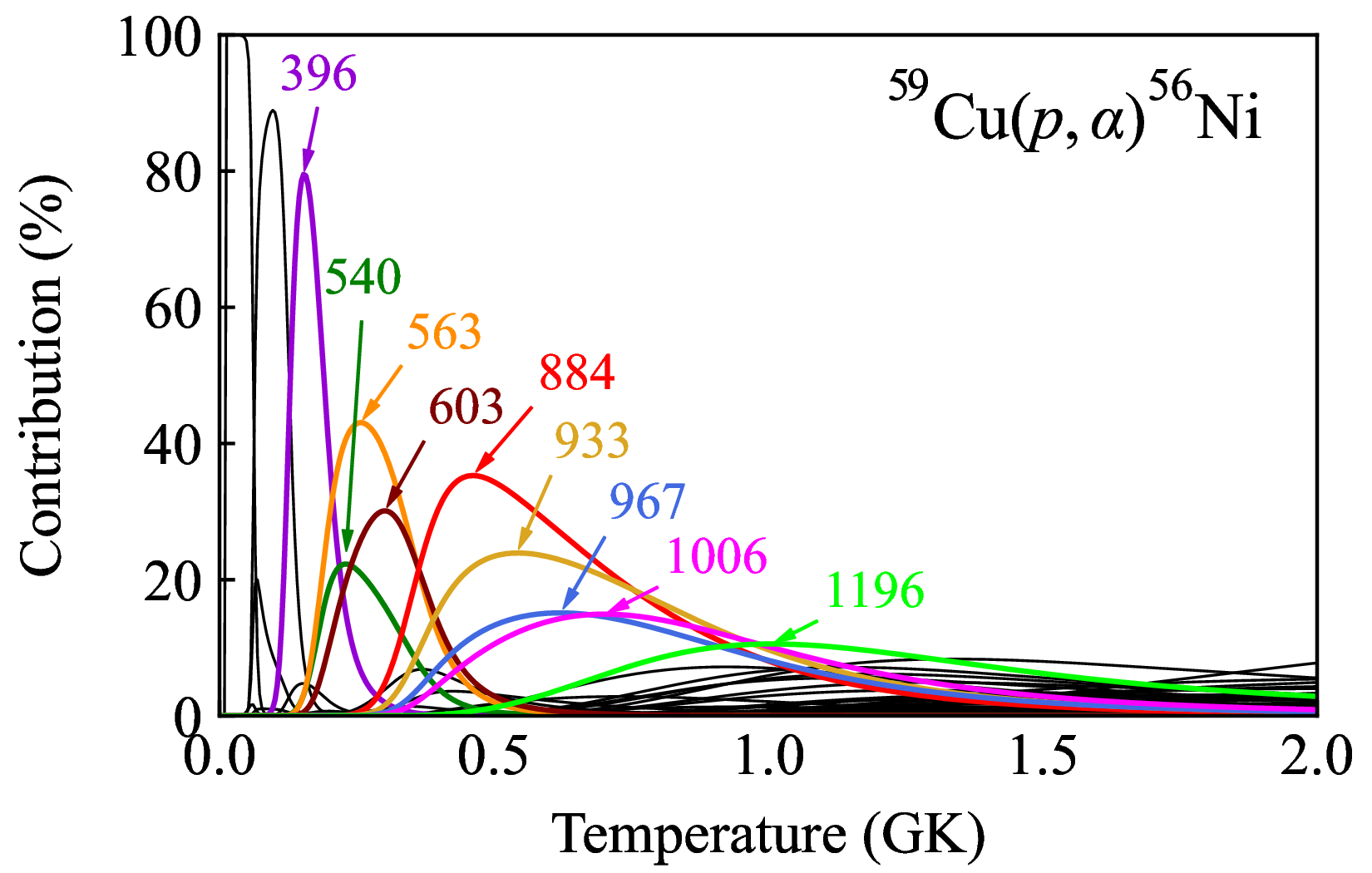}
\caption{\label{Rate_Contribution} Fractional contributions of 128 shell-model predicted resonances to the $^{59}$Cu$(p,\gamma)^{60}$Zn (upper) and $^{59}$Cu$(p,\alpha)^{56}$Ni (lower) reaction rates. The most influential resonances are labeled with their corresponding resonance energies in keV.}
\end{center}
\end{figure}

To enhance our understanding of $^{60}$Ga decay properties, we performed shell-model calculations in the truncated $fp$-shell model space with the GPFX1A Hamiltonian~\cite{Honma_PRC2004} using the \textsc{NuShellX@MSU} code~\cite{Brown_NDS2014}. The newly evaluated $^{60}$Ga $Q_\mathrm{EC}=14160(15)$~keV~\cite{Orrigo_PRC2021,Paul_PRC2021,Wang_PRL2023} was incorporated into the calculation. We obtained 900 $^{60}$Zn states populated by $^{60}$Ga decay up to $E_x=12.6$~MeV, with 300 states each for $J^\pi=1^+,2^+,3^+$. A quenching factor $q^2=0.6$ for the matrix elements of the Gamow-Teller operator was used to calculate the $\beta$ feedings in $^{60}$Ga decay. We calculated the decay widths $\Gamma_\gamma$ and $\Gamma_p$ for 128 resonances with $J^\pi=0^+,1^+,2^+,3^+,4^+,5^+$ up to $E_x=7.3$~MeV, corresponding to the upper end of the $^{59}$Cu$(p,\alpha)$ Gamow window at 1.5~GK. We also calculated the average decay widths $\Gamma_\gamma$, $\Gamma_p$, and $\Gamma_\alpha$ using the statistical model code \textsc{smaragd}~\cite{Rauscher_IJMPE2011,Rauscher_SMARAGD2014}. We adopted the shell-model calculated $\Gamma_\gamma$ and $\Gamma_p$ and the statistical-model calculated $\Gamma_\alpha$ to calculate the $^{59}$Cu$(p,\gamma)^{60}$Zn and $^{59}$Cu$(p,\alpha)^{56}$Ni reaction rates by combining all 128 positive parity resonances. The fractional contributions of each resonance are shown in Fig.~\ref{Rate_Contribution}. The statistical model calculation indicates that the level densities for $1^-$ and $2^-$ states in $^{60}$Zn fall below 1~MeV$^{-1}$ at excitation energies of 7.2 and 6.9~MeV, respectively. This suggests that $\ell=0$ resonances are less likely to be present within the Gamow window and to significantly contribute to the total reaction rate. Table~\ref{Key_resonances} summarizes the properties of the six most influential $^{59}$Cu$(p,\gamma)^{60}$Zn $\ell=1$ resonances. It should be noted that the uncertainties of the excitation/resonance energies are on the order of 200~keV. The resonances listed in Table~\ref{Key_resonances} are not necessarily the specific resonances that our experiment aims to identify but rather represent a typical potential scenario that we may encounter. As realistic $\Gamma_\alpha$ values vary much more than the average $\Gamma_\alpha$ calculated by the statistical model, it is likely that the influential $^{59}$Cu$(p,\alpha)^{56}$Ni resonances are fewer than those labeled in the lower panel of Fig.~\ref{Rate_Contribution}. Any $^{60}$Zn resonances that we are able to discover through $^{60}$Ga $\beta$ decay will provide valuable experimental constraints on the $^{59}$Cu$(p,\gamma)^{60}$Zn and $^{59}$Cu$(p,\alpha)^{56}$Ni reaction rates.

A theoretical reaction rate calculation with principled uncertainty quantification will be discussed in a forthcoming paper~\cite{Adams_PRC2025}, in which all the nuclear physics properties entering into the reaction rate calculation will be sampled according to appropriate probability density functions~\cite{Longland_NPA2010_1,Iliadis_NPA2010_2,Iliadis_NPA2010_3,Mohr_PRC2014,Iliadis_JPG2015}.

\FloatBarrier
\section{Simulations}

\begin{figure}[htbp]
\begin{center}
\includegraphics[width=8.6cm]{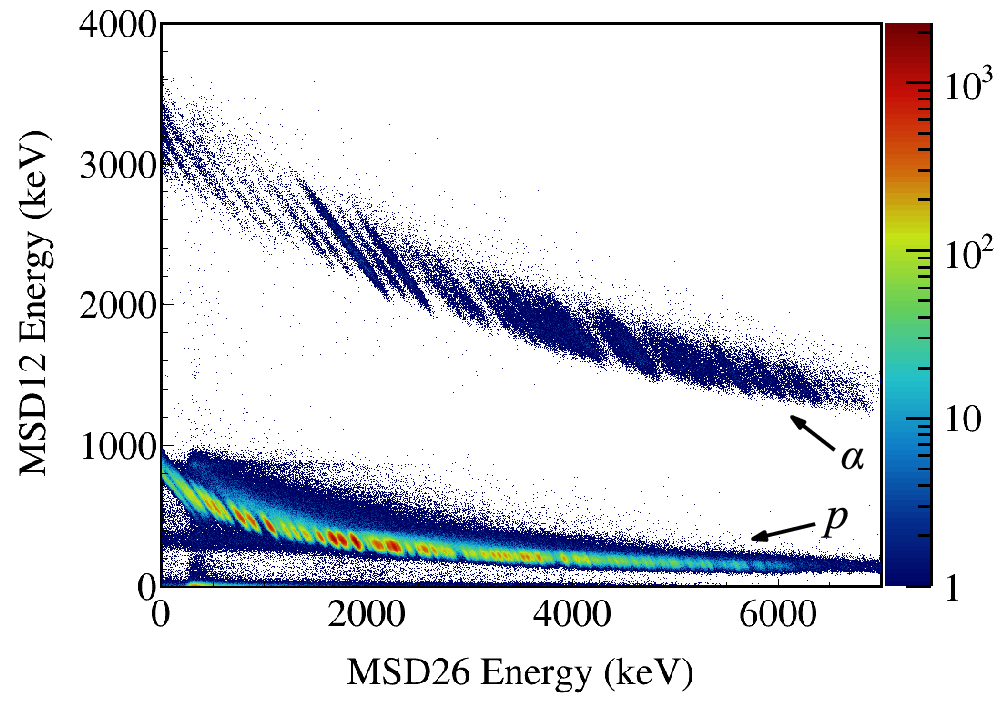}
\includegraphics[width=8.6cm]{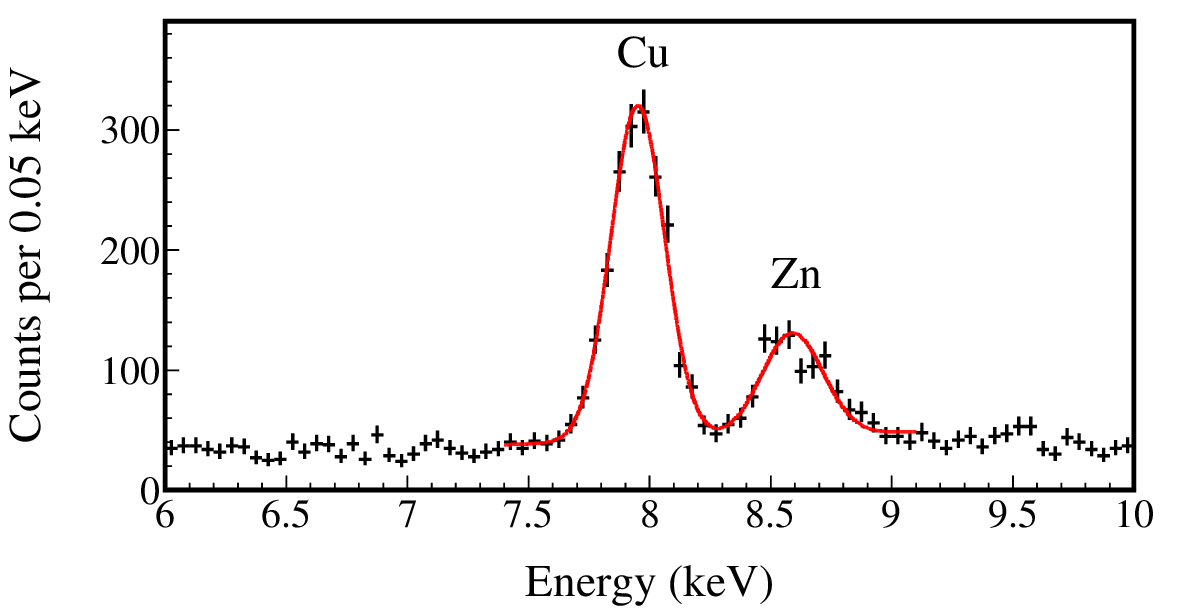}
\includegraphics[width=8.6cm]{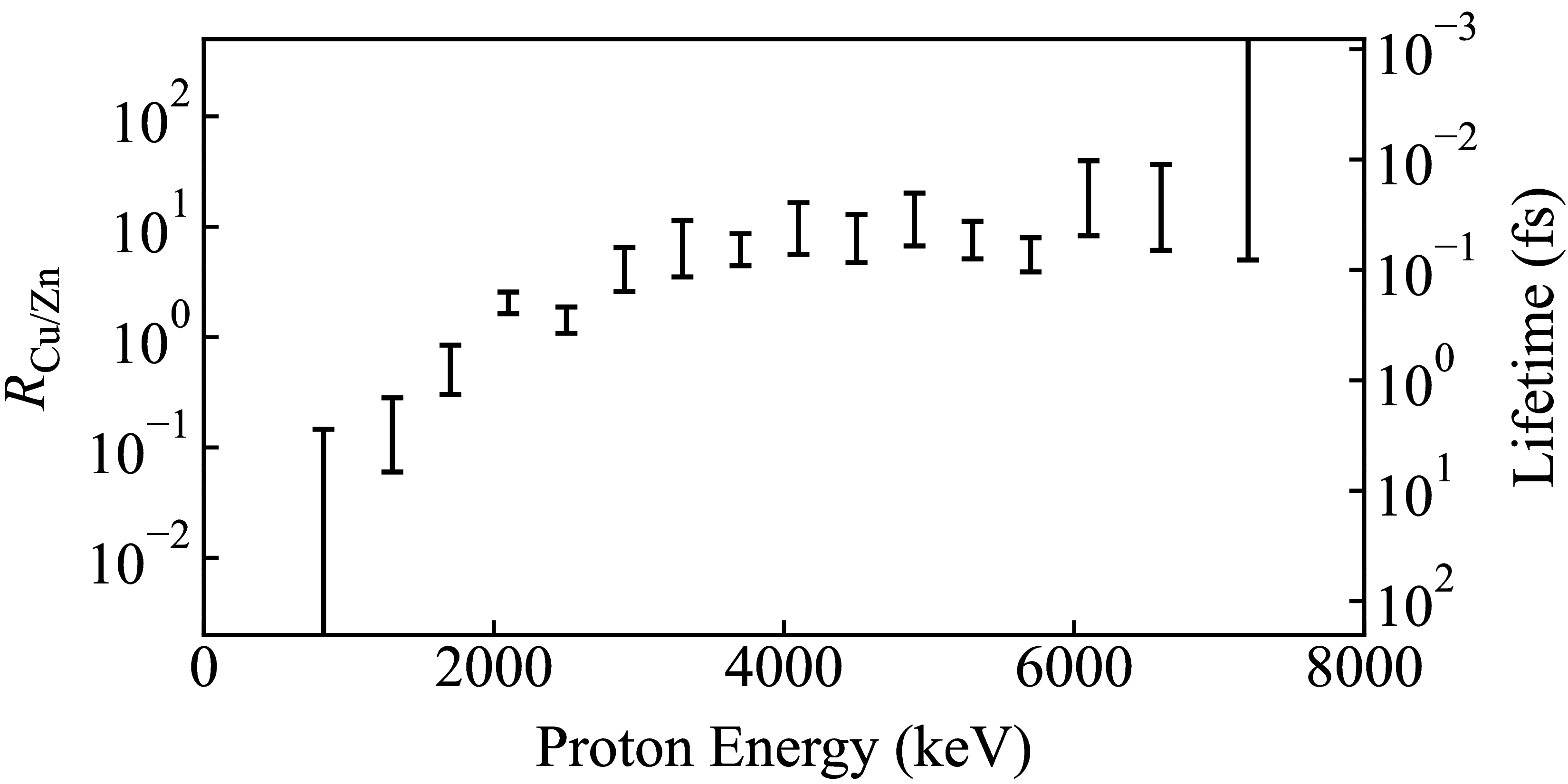}
\caption{\label{px_coin_spec_60Ga} Top panel: charged-particle $\Delta E$-$E$ spectrum simulated by incorporating the theoretical decay properties of $^{60}$Ga and the measured detector responses. Middle panel: X-ray spectrum gated by all protons in the $\Delta E$-$E$ spectrum, yielding a total X-ray ratio of $R_{\text{Cu/Zn}} = 3.2(3)$. A double-Gaussian with a linear background fit is superimposed on the Cu and Zn $K_\alpha$ peaks. Bottom panel: Cu/Zn $K_\alpha$ X-ray count ratio and the inferred lifetime as a function of coincident proton energy. The error bars represent statistical uncertainties, where the leftmost and rightmost error bars indicate upper or lower limits.}
\end{center}
\end{figure}

To assess the feasibility of the $^{60}$Ga decay measurement with LIBRA, we have developed \textsc{geant}4 simulations incorporating the theoretical $^{60}$Ga decay scheme and the known decay schemes of the daughter nuclei~\cite{Huo_NDS2011_A56,Basunia_NDS2018_A59,Browne_NDS2013_A60}, the detector responses characterized based on radioactive source tests, and the integrated $^{60}$Ga beam intensity over a six-day period ($1.5\times10^{10}$). The resulting simulated proton and $\alpha$ particle identification spectrum by the $\Delta E$-$E$ telescope and the proton-gated X-ray spectrum by LEGe are shown in Fig.~\ref{px_coin_spec_60Ga}. As demonstrated in the charged-particle $\Delta E$-$E$ spectrum, our setup enables clear separation of most proton and $\alpha$ emission branches, allowing for accurate determination of their respective decay branching ratios.

One of the key observables offered by LIBRA is the proton-X-ray coincidence. The Cu/Zn $K_\alpha$ X-ray count ratio can be determined by integrating the 8.0- and 8.6-keV X-ray peaks observed in coincidence with protons. The Zn $K_\alpha$ radiative transition probability is 41.4\%, compared to 38.7\% for Cu~\cite{Perkins_EADL1991}, and the LEGe detection efficiency for 8.6-keV photons is 7.8\%, compared to 7.4\% at 8.0~keV~(Fig.~\ref{LEGe_XEfficiency}). Consequently, we need to apply two correction factors of $F=1.07$ for fluorescence yields and $E=1.05$ for efficiencies when extracting the lifetime of the proton-emitting state in $^{60}$Zn from the observed Cu/Zn $K_\alpha$ X-ray count ratio:

\begin{align}
\tau_{p-\text{emit}} &= \frac{\tau_{K\text{shell(Zn)}}}{R_{\text{Cu/Zn}}}, \\
R_{\text{Cu/Zn}} &= \frac{I_{K_{\alpha}\text{(Cu)}} \times F \times E}{I_{K_{\alpha}\text{(Zn)}}},
\label{eq:ZnCuRatio}
\end{align}

The bottom panel of Fig.~\ref{px_coin_spec_60Ga} shows the Cu/Zn $K_\alpha$ X-ray count ratios as a function of coincident proton energies, along with the corresponding lifetimes of proton-emitting states in $^{60}$Zn. Only the X-ray count statistical uncertainty is taken into account. The integrated X-ray ratio of $R_{\text{Cu/Zn}}=3.2(3)$ corresponds to a lifetime $\tau_{p-\text{emit}}=0.126(11)$~fs, which is an average for all $^{60}$Zn proton-emitting states. The main source of systematic uncertainty is the recommended Zn $K$-shell vacancy width $\Gamma_{K\text{shell(Zn)}}=1.62$~eV~\cite{Campbell_ADNDT2001}, adjusted based on the calculated $\Gamma_{K\text{shell(Zn)}}=1.56$~eV from Ref.~\cite{Perkins_EADL1991}. A resonant Raman scattering measurement reported $\Gamma_{K\text{shell(Zn)}}=1.9(1)$~eV~\cite{Hamalainen_JPCM1989}, which is consistent with the recommended value, considering the estimated uncertainty of 5$-$25\% for atomic numbers below 30~\cite{Campbell_ADNDT2001}.

Notably, the nuclear lifetimes can be reliably determined within a sensitivity range that spans approximately one order of magnitude above and below the Zn $K$-shell vacancy lifetime of 0.4~fs~(Table~\ref{PXCT_history}). This timescale is typical for most nuclear resonances but extremely challenging for conventional lifetime measurement techniques. Should the experimental statistics fall short of the current estimates, we may obtain only upper or lower limits on certain branches, which would still be meaningful for constraining the astrophysical reaction rates.

\section{Summary \& Outlook}
We present the design, construction, simulation, and radioactive source testing of LIBRA. The system is capable of detecting all types of charged particles and photons emitted in EC/$\beta^+$ decay, which will enable us to measure the energies, proton-, $\alpha$-, and $\gamma$-decay branching ratios of resonances. We will also utilize PXCT to determine the lifetimes of resonances populated by EC. Proton/$\alpha$-$\gamma$ coincidences provide information on the proton/$\alpha$-emitting states in the compound nucleus and the ground and excited states of daughter nuclei, pertinent to both the entrance and exit channels for particle-induced reactions. LIBRA data can also provide nuclear level densities and transmission coefficients needed for calculating reaction rates using the Hauser-Feshbach statistical model.

Utilizing LIBRA for $^{60}$Ga EC/$\beta^+$ decay offers the unique advantage of obtaining a comprehensive set of nuclear data in a single experiment. This capability efficiently addresses a key limitation of traditional indirect methods that they often yield only partial necessary nuclear data and thus require multiple experiments. By combining the nuclear data acquired through LIBRA, we may provide experimentally constrained thermonuclear rates of the $^{59}$Cu$(p,\gamma)^{60}$Zn and $^{59}$Cu$(p,\alpha)^{56}$Ni reactions to XRB models. The results could help us better understand the NiCu cycle, reduce the nuclear physics uncertainties in modeling the light curves and nucleosynthesis of XRBs, and ultimately facilitate the comparisons between model predictions and astronomical observations.

LIBRA holds the potential for studying other important reaction rates in the $rp$ process. As shown in Fig.~\ref{NiCu_Cycle}, $^{64}$Ge plays an analogous role in the ZnGa cycle to that of $^{60}$Zn in the NiCu cycle~\cite{Cyburt_APJ2016,Meisel_APJ2019,Lu_PRC2024}. A notable difference is that the allowed $\beta$ transitions of the $0^+$ $^{64}$As ground state populate the $0^+$ and $1^+$ states in $^{64}$Ge~\cite{Singh_NDS2021_A64}. Given the comparable $Q_{\mathrm{EC}}$, half-lives, proton/$\alpha$-separation energies, and key X-ray energies (Table~\ref{PXCT_history}), it is technically feasible to utilize LIBRA for the $^{64}$As EC/$\beta^+$ decay experiment to address the competition between the $^{63}$Ga$(p,\gamma)^{64}$Ge and $^{63}$Ga$(p,\alpha)^{60}$Zn reactions.


\section{Acknowledgments}
We gratefully acknowledge Stephen Gillespie, Aaron Chester, Giordano Cerizza, Craig Snow, and the FRIB Mechanical Engineering Department for their technical support. We would like to thank John Hardy, Tibor  Kib\'{e}di, and Daid Kahl for the helpful discussions. This work was supported by the U.S. National Science Foundation under Grants Nos. PHY-1913554, PHY-2110365, and PHY-2209429, and the U.S. Department of Energy, Office of Science, under Awards Nos. DE-SC0016052 and DE-SC0023529.

\end{document}